\documentclass{article}

\usepackage{amsmath}
\usepackage{arxiv}

\usepackage[utf8]{inputenc} 
\usepackage[T1]{fontenc}    
\usepackage{hyperref}       
\usepackage{url}            
\usepackage{booktabs}       
\usepackage{amsfonts}       
\usepackage{nicefrac}       
\usepackage{microtype}      
\usepackage{lipsum}
\usepackage{graphicx}
\graphicspath{ {./images/} }

\title{Climate Vulnerability and Community Health: Identifying Greensboro Neighborhoods at Intersectional Risk}

\author{
Rehinatu Usman \\
Department of Earth, Environment and Planning\\
East Carolina University\\
Greenville, North Carolina, USA\\
\texttt{rehinatusman@gmail.com}
\And
Onyedikachi J. Okeke \\
Department of Geography and Environmental Studies\\
College of Liberal Arts\\
Texas State University\\
San Marcos, TX, USA
}

\begin{document}

\maketitle

\begin{abstract}
This study develops an integrated, intersectional climate vulnerability assessment for Greensboro, North Carolina, a midsize city in the rapidly changing American Southeast. Moving beyond generalized mapping, we combine demographic, socioeconomic, health, and environmental data at the census tract level to identify neighborhoods where flood exposure, chronic health burdens, and social disadvantage spatially converge. Through k-means and hierarchical clustering, we identify four distinct neighborhood typologies, including a critically high-risk cluster characterized by high flood exposure, extreme poverty, poor respiratory health, and aging housing. The findings demonstrate that climate-related risks are not randomly distributed but systematically cluster in historically marginalized communities, revealing a clear environmental justice disparity. This place-based typology approach provides a targeted framework for policymakers to design integrated interventions that bridge flood management, public health, housing, and social services to build equitable urban resilience
\end{abstract}


\section{Introduction}

Climate change is no longer a distant threat but a present-day determinant of urban health, manifesting through intensifying heatwaves, more severe flooding, and degraded air quality \cite{EPA2023}. These hazards create a public health crisis whose burdens are inequitably distributed, systematically following the contours of historical injustice and contemporary socioeconomic disparity. In cities across the United States, from Phoenix to Philadelphia, the neighborhoods facing the highest exposure to environmental risk are consistently those marginalized by race, class, and a legacy of discriminatory policy \cite{Harlan2012, May2023, Cutter2003, Chakraborty2019}. Greensboro, North Carolina, a mid-sized city with a deep history of industrialization and segregation, exemplifies this national pattern. Here, the urgent challenge for public health and urban planning is to move from generalized awareness to spatially precise identification of communities at intersectional risk. This study argues that the neighborhoods most vulnerable to climate-related health harms in Greensboro can be accurately pinpointed through a layered analysis that overlaps contemporary socioeconomic vulnerability, environmental exposure, and the historical geography of redlining. Such an approach is critical for transforming broad equity goals into targeted, effective interventions that address both current threats and their root causes.

Extensive research has established the exposure-sensitivity-adaptive capacity framework as the dominant paradigm for assessing climate vulnerability \cite{Yu2021, Reid2008, Cutter2003}. Studies in diverse geographic contexts demonstrate that vulnerability is not random but spatially clustered. Research in Maricopa County, Arizona, linked higher rates of heat-related deaths to neighborhoods with lower socioeconomic status, higher proportions of elderly or socially isolated residents, and less vegetation \cite{Harlan2012, Harlan2006, Voelkel2018}. Similarly, a national analysis revealed high heterogeneity in vulnerability at the census tract level, with significant clustering of social and environmental risk factors \cite{Lewis2023}. A critical advancement in this literature is the explicit connection between historical racist housing policies, such as 1930s-era redlining by the Home Owners' Loan Corporation, and present-day environmental risk. Studies in Philadelphia and other cities have shown that areas historically graded as hazardous or definitely declining remain hotspots for extreme heat, poor air quality, and sparse canopy cover decades later \cite{May2023, Breakey2024, Chakraborty2019}. However, a significant gap remains in applying this integrated, historically informed vulnerability assessment to mid-sized cities in the American Southeast, a region experiencing rapid climatic changes. Many existing models are derived from large coastal or southwestern metropolitan areas, leaving a need for fine-grained, context-specific analyses for cities like Greensboro, where the intersection of climate risk with specific histories of racial and economic segregation requires direct examination \cite{Anderko2014, Adepoju2021}.

The primary beneficiaries of this research are municipal agencies, public health officials, urban planners, and community-based organizations in Greensboro. For agencies such as the Greensboro Department of Health and Human Services or the Office of Sustainability and Resilience, these findings provide an evidence-based spatial blueprint for prioritizing investments. Resources for tree planting, cooling center placement, floodplain buyouts, and health outreach can be directed with greater precision to the census tracts where need is most acute. The methodology, while tailored to Greensboro’s unique demographic and historical context, offers a transferable model for peer cities such as Winston-Salem, Durham, or Chattanooga. The core analytical process of integrating high-resolution social, environmental, and historical spatial data is broadly applicable. Its emphasis on redlining as a structural determinant ensures relevance to the many U.S. cities whose contemporary landscapes were shaped by these discriminatory maps, positioning the approach as a tool for advancing climate justice beyond a single case study.

This analysis is built upon a reproducible, open-source geospatial workflow to ensure transparency, accessibility, and potential for replication. Data are synthesized from multiple public sources, including U.S. Census Bureau American Community Survey estimates for socioeconomic and demographic variables, satellite-derived metrics such as the Normalized Difference Vegetation Index to quantify tree canopy, Landsat thermal data to calculate land surface temperature, and digitized historical Home Owners' Loan Corporation redlining maps from the University of Richmond Mapping Inequality project. Integration and analysis are conducted using Python-based geospatial libraries, including GeoPandas and Rasterio, within Jupyter notebooks \cite{USCensusACS2023, Jordahl2020, Gillies2013, Kluyver2016, MappingInequality}. This commitment to open science supports academic replication and provides a practical template for local governments and advocacy groups with limited technical resources. Final outputs include static and interactive maps designed to communicate complex vulnerability scores clearly to policymakers and community stakeholders.

The central aim of this research is to develop and apply an intersectional climate vulnerability index for Greensboro that identifies neighborhoods at disproportionate risk for climate-related health impacts. This aim is operationalized through three objectives. First, to construct a composite index by statistically integrating validated indicators of social sensitivity, such as poverty and age over 65, adaptive capacity, such as access to a vehicle and housing age and quality, and environmental exposure, such as low vegetation cover and high impervious surface area. Second, to analyze the spatial correlation between contemporary high-vulnerability zones and the historical boundaries of redlined neighborhoods in Greensboro. Third, to characterize the demographic, socioeconomic, and environmental profiles of census tracts in the highest decile of intersectional risk, producing a prioritized list for intervention.

This study demonstrates that the most climate-vulnerable neighborhoods in Greensboro are those where low median household income, high proportions of African American residents, sparse tree canopy, and historical redlining designation spatially converge. The significance of this finding lies in its actionable specificity, translating the concept of climate injustice into a precise, mappable reality for local decision makers. The research is guided by two questions: (1) How do indicators of social vulnerability and environmental risk intersect at the census tract level to create zones of compounded climate health risk in Greensboro? (2) To what extent do these zones of highest intersectional risk align with neighborhoods historically redlined as hazardous or definitely declining? By addressing these questions, the study links past institutional discrimination to present-day vulnerability and informs equitable resilience strategies that seek to repair as well as protect. The paper proceeds with a review of relevant literature, a detailed methodology, a presentation of results including vulnerability maps, and a discussion of implications for policy, planning, and community health in Greensboro.

\section{Methods and Materials}

\subsection{Study Area and Geographic Scope}

This study focused on Greensboro, North Carolina, a mid-sized city within Guilford County characterized by diverse socioeconomic conditions and environmental exposures. The analysis was conducted at the census tract level ($n=94$), representing the finest geographic scale at which comprehensive health, demographic, and environmental data converge. Census tracts were selected because they typically encompass between 1,200 and 8,000 residents and are designed to be relatively homogeneous with respect to population characteristics and living conditions \cite{USCensusGEOID2020}. The study area was defined by intersecting 2024 TIGER/Line census tract boundaries with the municipal boundary of Greensboro, ensuring that the analysis captured neighborhoods within the city’s jurisdiction while maintaining consistency with current administrative geography \cite{TIGERLine2024}.

The selection of Greensboro as a case study is particularly relevant for climate vulnerability research due to its location within the Piedmont region of North Carolina, which experiences both fluvial and pluvial flooding risks in combination with urban heat island effects and documented socioeconomic disparities in health outcomes \cite{Gaither2021}. The city’s demographic composition, approximately 48\% White, 41\% Black, and 8\% Hispanic or Latino residents, provides an important context for examining environmental justice dimensions of climate vulnerability \cite{USCensusACS2023}.

\subsection{Data Sources and Integration Framework}

The methodological approach employed a multi-source data integration framework, drawing from federal, state, and local datasets to construct a comprehensive vulnerability assessment. All datasets were standardized to the 2020 census tract geography using Federal Information Processing System codes, with careful attention to temporal alignment to ensure comparability across sources.

\subsubsection{Demographic and Socioeconomic Data}

Demographic and housing characteristics were obtained from the 2019--2023 American Community Survey five-year estimates, which provide reliable small-area statistics for socioeconomic indicators \cite{USCensusACS2023, Spielman2014}. Data were accessed via the U.S. Census Bureau Application Programming Interface using a custom Python workflow that automated variable retrieval and error handling \cite{USCensusACS2023}. Key variables extracted included housing tenure and occupancy, year structure built, poverty status, educational attainment, disability status, vehicle availability, and limited English proficiency.

All variables were transformed into percentage rates by dividing subgroup counts by appropriate denominator populations. For example, poverty rate was calculated as the population below the poverty line divided by the total poverty universe. This normalization controlled for population size differences across census tracts and enabled comparative spatial analysis.

\subsubsection{Health Outcome Data}

Health indicators were sourced from the Centers for Disease Control and Prevention PLACES 2024 dataset, which provides census tract-level estimates for population health outcomes \cite{CDCPLACES2024}. PLACES data employ a validated small-area estimation methodology that integrates Behavioral Risk Factor Surveillance System data with American Community Survey demographics through multilevel regression and poststratification techniques \cite{Zhang2021}. This study focused on five respiratory and cardiovascular health measures most relevant to climate vulnerability: current asthma prevalence among adults, chronic obstructive pulmonary disease prevalence, coronary heart disease prevalence, poor or fair self-rated health status, and prevalence of any disability among adults.

\subsubsection{Environmental Exposure Data}

Environmental exposure indicators were operationalized using multiple datasets representing distinct dimensions of climate risk. Flood exposure was derived from the Federal Emergency Management Agency National Flood Hazard Layer, accessed through the ArcGIS REST Application Programming Interface \cite{FEMANFHL2024}. The dataset delineates Special Flood Hazard Areas corresponding to the one percent annual chance floodplain. A spatial overlay was conducted using a one-meter buffer around flood zone boundaries to create continuous polygons, followed by an intersection with census tract boundaries to compute the proportion of each tract located within flood hazard zones. Flood exposure was calculated as

\begin{equation}
\text{Flood Exposure Index}_i =
\frac{A_i^{\text{flood}}}{A_i^{\text{total}}} \times 100,
\end{equation}

where $A_i^{\text{flood}}$ represents the area of census tract $i$ intersecting flood hazard zones and $A_i^{\text{total}}$ denotes the total area of tract $i$.

Air quality conditions were represented using county-level annual Air Quality Index values for 2023 obtained from the U.S. Environmental Protection Agency AirData system \cite{EPA2023}. Although tract-level air pollution measures would provide higher spatial resolution, county-level Air Quality Index values offer a proxy for regional air quality patterns that influence respiratory health outcomes \cite{Miranda2011}.

\subsection{Geographic and Administrative Data}

Census tract boundaries were obtained from the U.S. Census Bureau TIGER/Line 2024 dataset, representing the most current administrative geography available \cite{TIGERLine2024}. All tract geometries were reprojected to a projected coordinate reference system (EPSG:3857) to ensure accurate area calculations and spatial operations. Municipal boundaries for Greensboro were similarly sourced from TIGER/Line place files and used to subset census tracts to those intersecting the city boundary.

\section{Data Processing and Quality Assurance}

A systematic data processing pipeline was implemented to ensure consistency, accuracy, and reproducibility throughout the analysis. All preprocessing steps were documented and automated using Python-based workflows to minimize manual intervention and reduce the potential for error.

\subsection{Data Cleaning and Transformation}

Initial data cleaning addressed common challenges associated with multi-source geospatial data integration.

\paragraph{Identifier Standardization}
All datasets were joined using a standardized 11-digit geographic identifier constructed from state (two digits), county (three digits), and census tract (six digits) Federal Information Processing System codes \cite{USCensusGEOID2020}. This approach ensured consistent matching across datasets despite differences in field naming conventions.

\paragraph{Outlier Treatment}
Extreme values were identified using Tukey’s fences method, defined as observations below $Q_1 - 1.5 \times IQR$ or above $Q_3 + 1.5 \times IQR$, where $Q_1$ and $Q_3$ denote the first and third quartiles and $IQR = Q_3 - Q_1$ \cite{Tukey1977}. Values identified as extreme were examined for potential data entry errors and, when confirmed as valid, were Winsorized at the 5th and 95th percentiles to reduce their influence on subsequent analyses.

\paragraph{Variable Transformation}
Continuous variables were evaluated for distributional properties using Shapiro--Wilk tests. Variables exhibiting substantial skewness, defined as absolute skewness greater than one, were log-transformed where theoretically appropriate to better approximate normality \cite{Osborne2010}.

\subsection{Composite Indicator Development}

To capture the multidimensional nature of climate vulnerability, composite indices were developed using a structured, multi-step procedure.

\paragraph{Normalization}
All component variables were normalized to a common 0--100 scale using min--max normalization:
\begin{equation}
x'_{ij} = \frac{x_{ij} - \min(x_j)}{\max(x_j) - \min(x_j)} \times 100,
\end{equation}
where $x_{ij}$ represents the value of variable $j$ for census tract $i$, and $\min(x_j)$ and $\max(x_j)$ denote the minimum and maximum values of variable $j$ across all tracts.

\paragraph{Health Vulnerability Score}
Five PLACES health indicators were aggregated using equal weights:
\begin{equation}
H_i = \frac{1}{5}\sum_{k=1}^{5} x'_{ik}.
\end{equation}

\paragraph{Socioeconomic Vulnerability Score}
Four American Community Survey socioeconomic indicators were similarly combined:
\begin{equation}
S_i = \frac{1}{4}\sum_{l=1}^{4} x'_{il}.
\end{equation}

\begin{figure}[htbp]
\centering
\setlength{\tabcolsep}{2pt}
\renewcommand{\arraystretch}{1}
\begin{tabular}{cccc}
\includegraphics[width=0.23\textwidth]{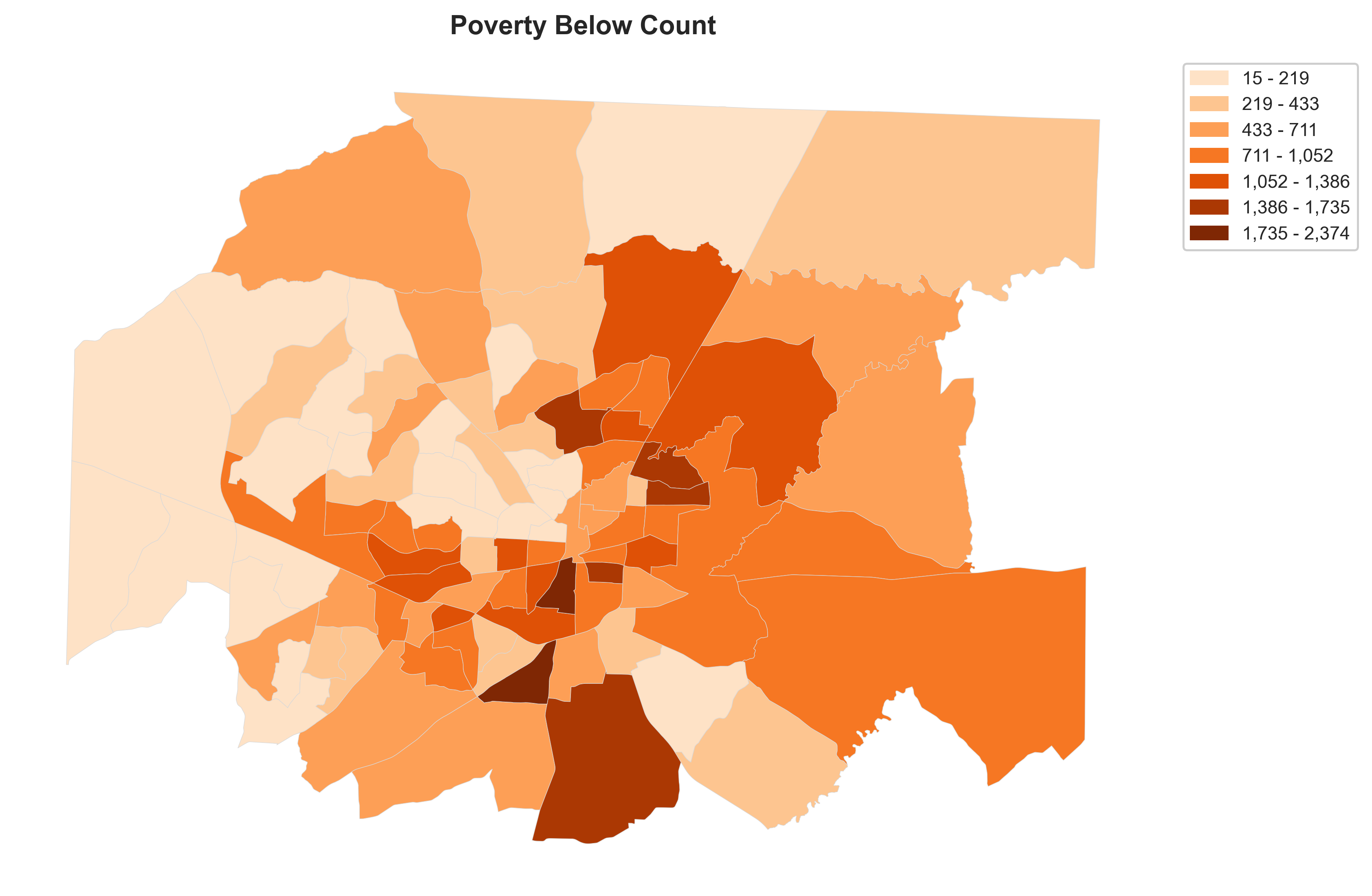} &
\includegraphics[width=0.23\textwidth]{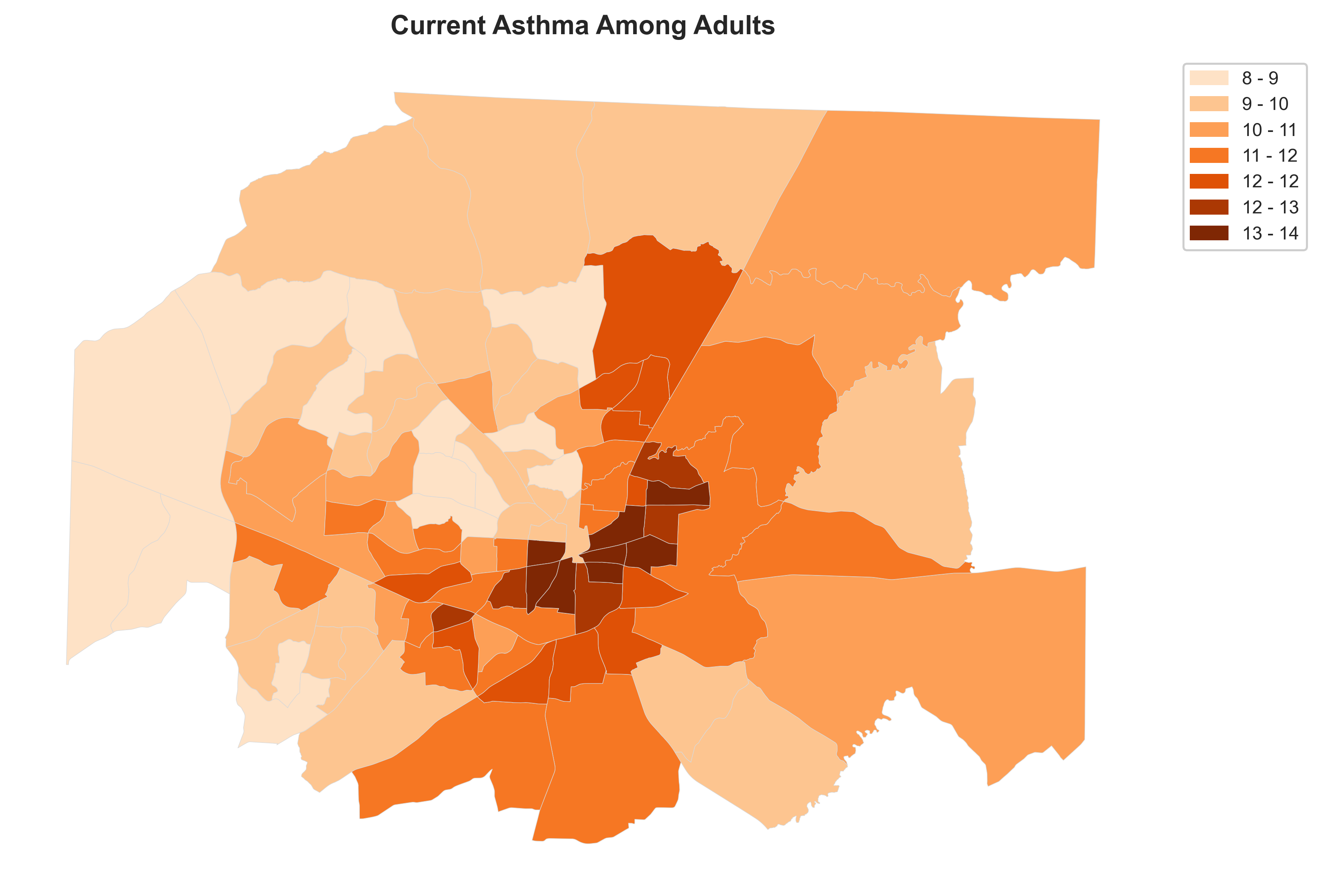} &
\includegraphics[width=0.23\textwidth]{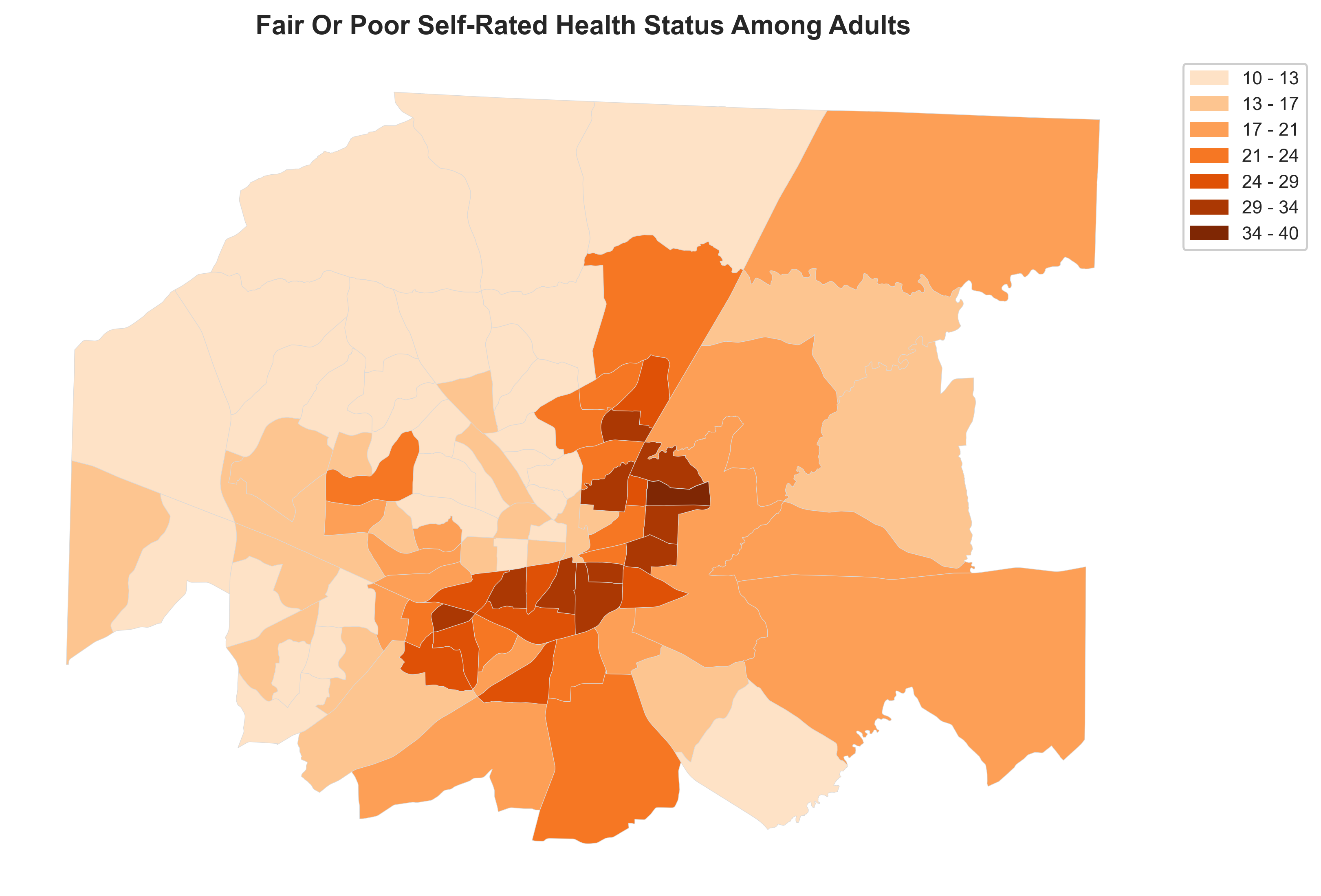} &
\includegraphics[width=0.23\textwidth]{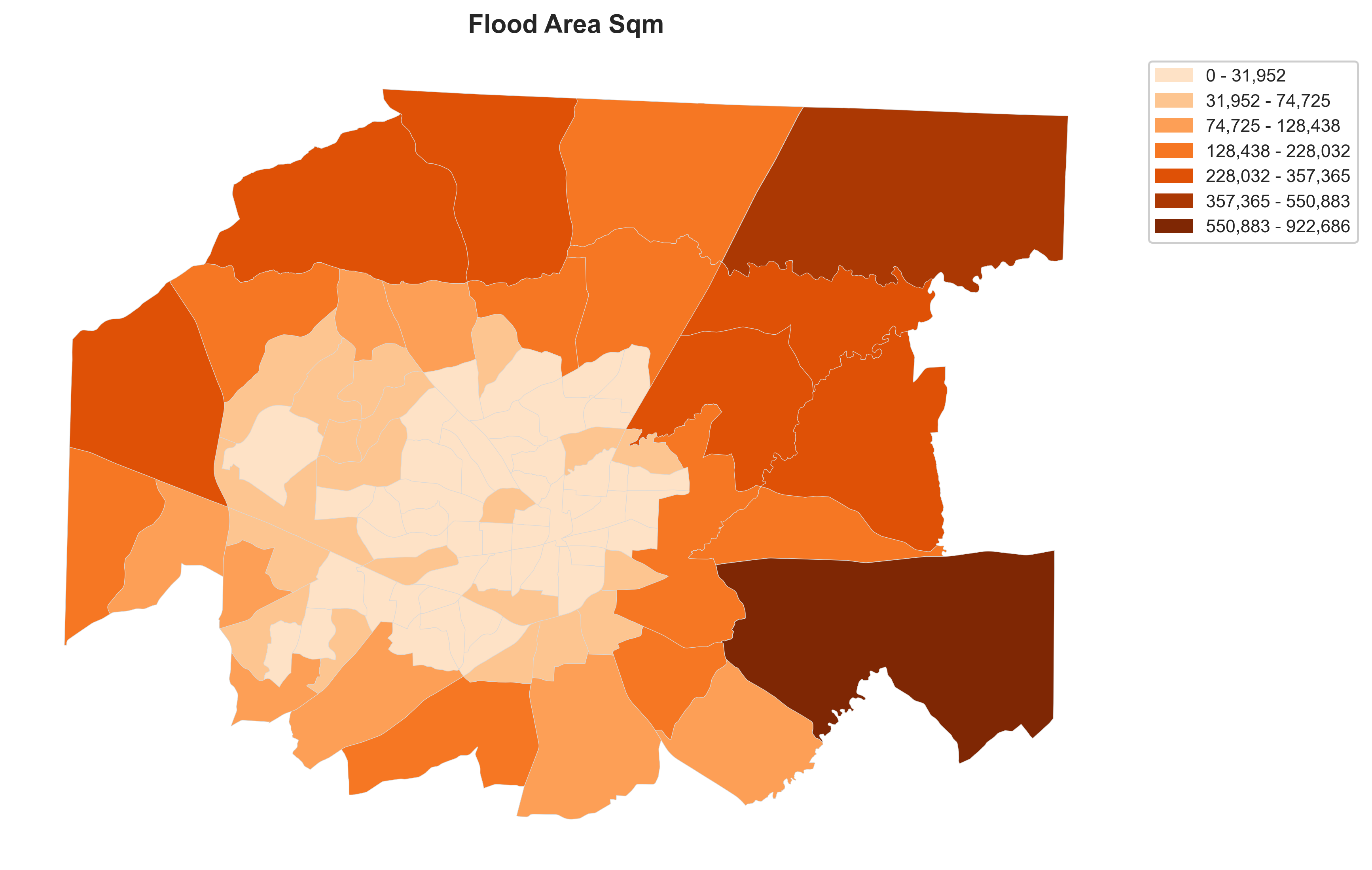} \\
\includegraphics[width=0.23\textwidth]{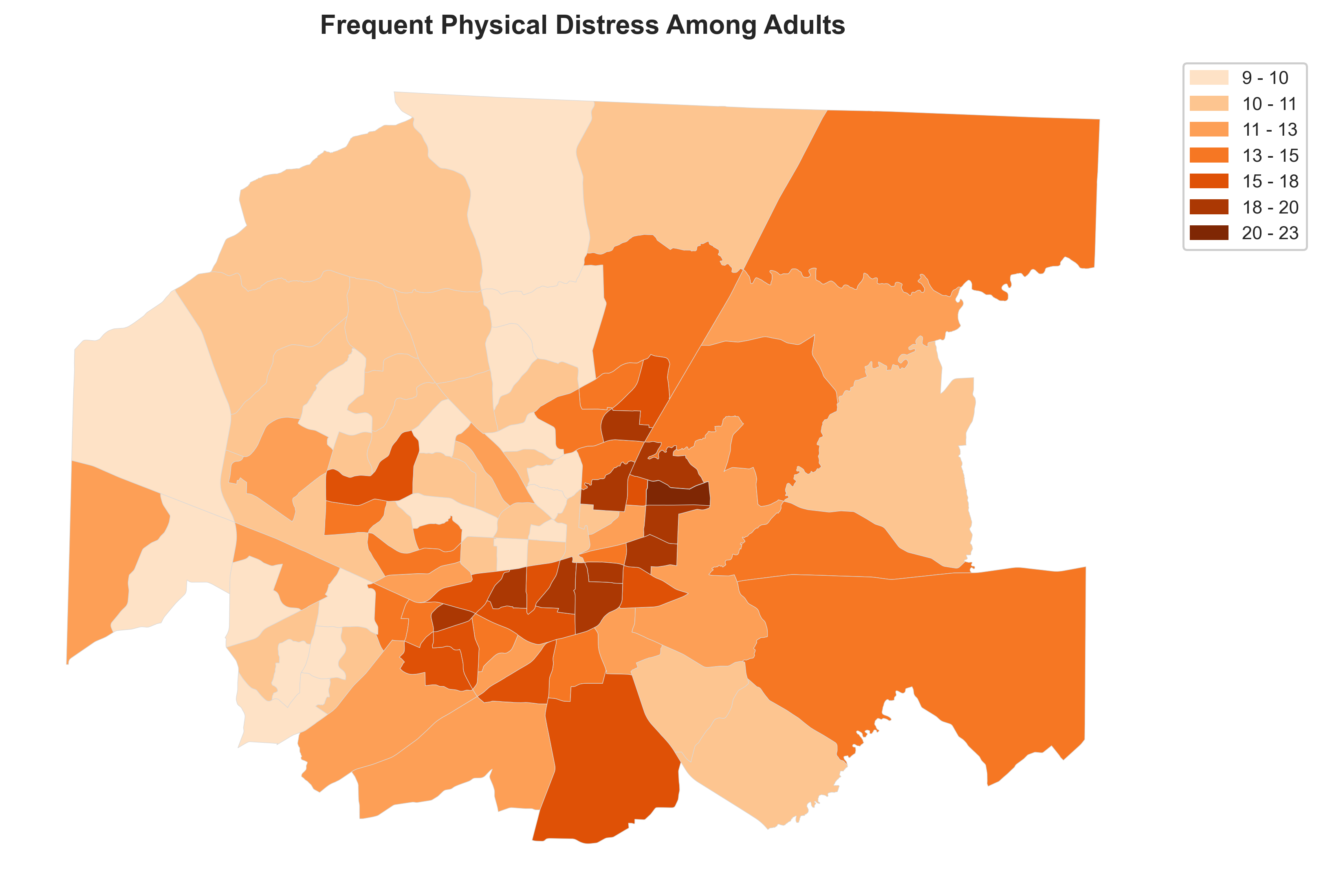} &
\includegraphics[width=0.23\textwidth]{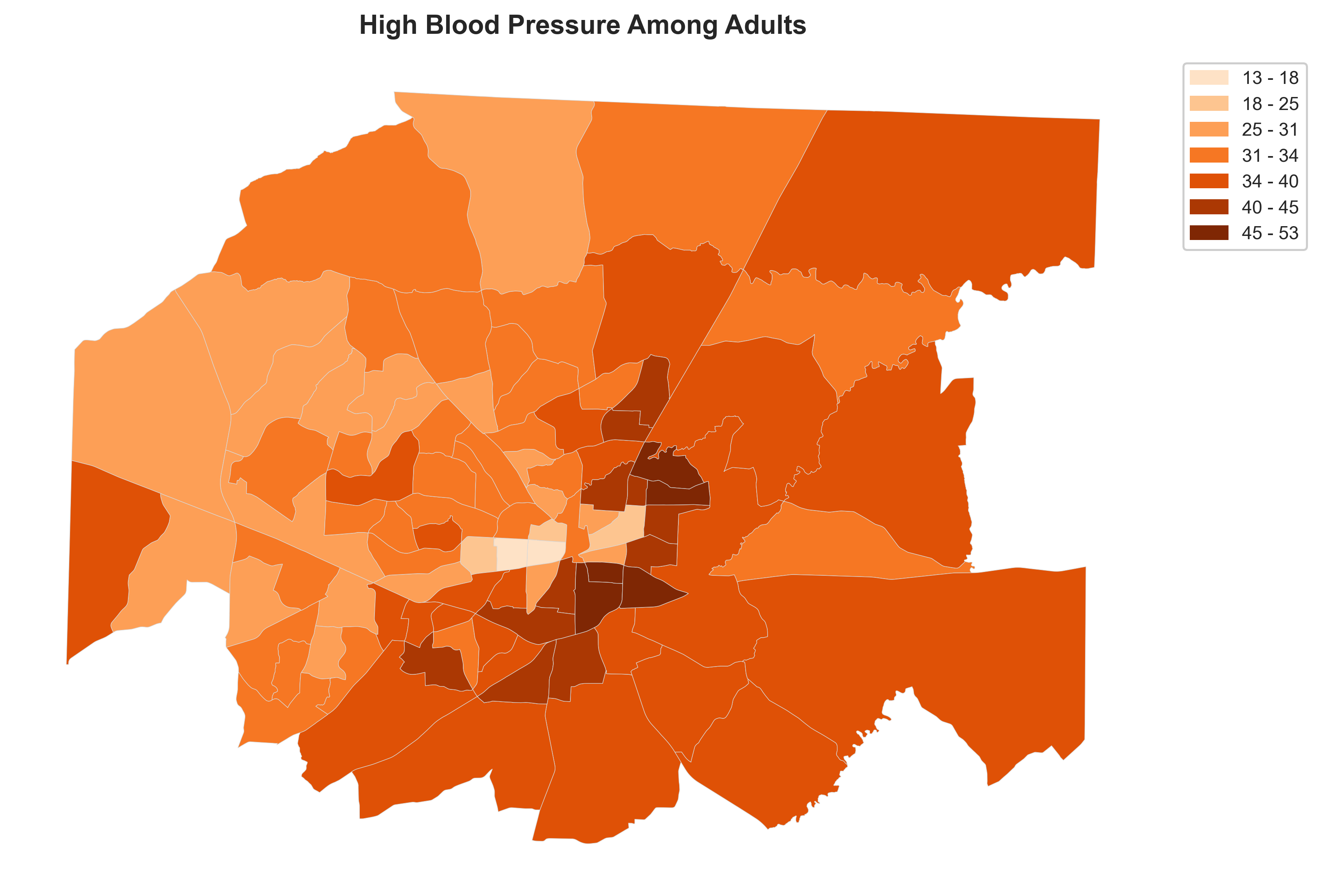} &
\includegraphics[width=0.23\textwidth]{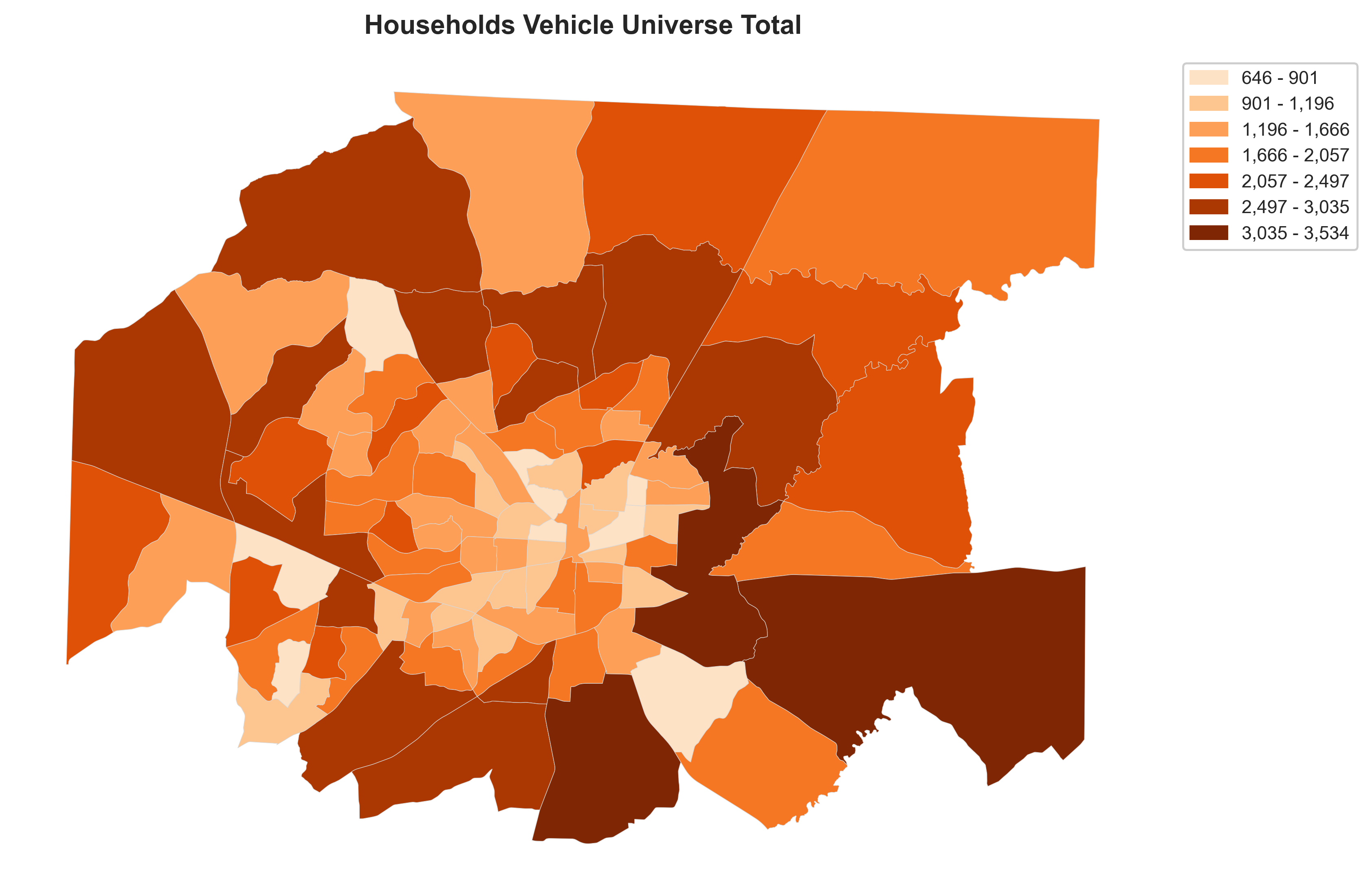} &
\includegraphics[width=0.23\textwidth]{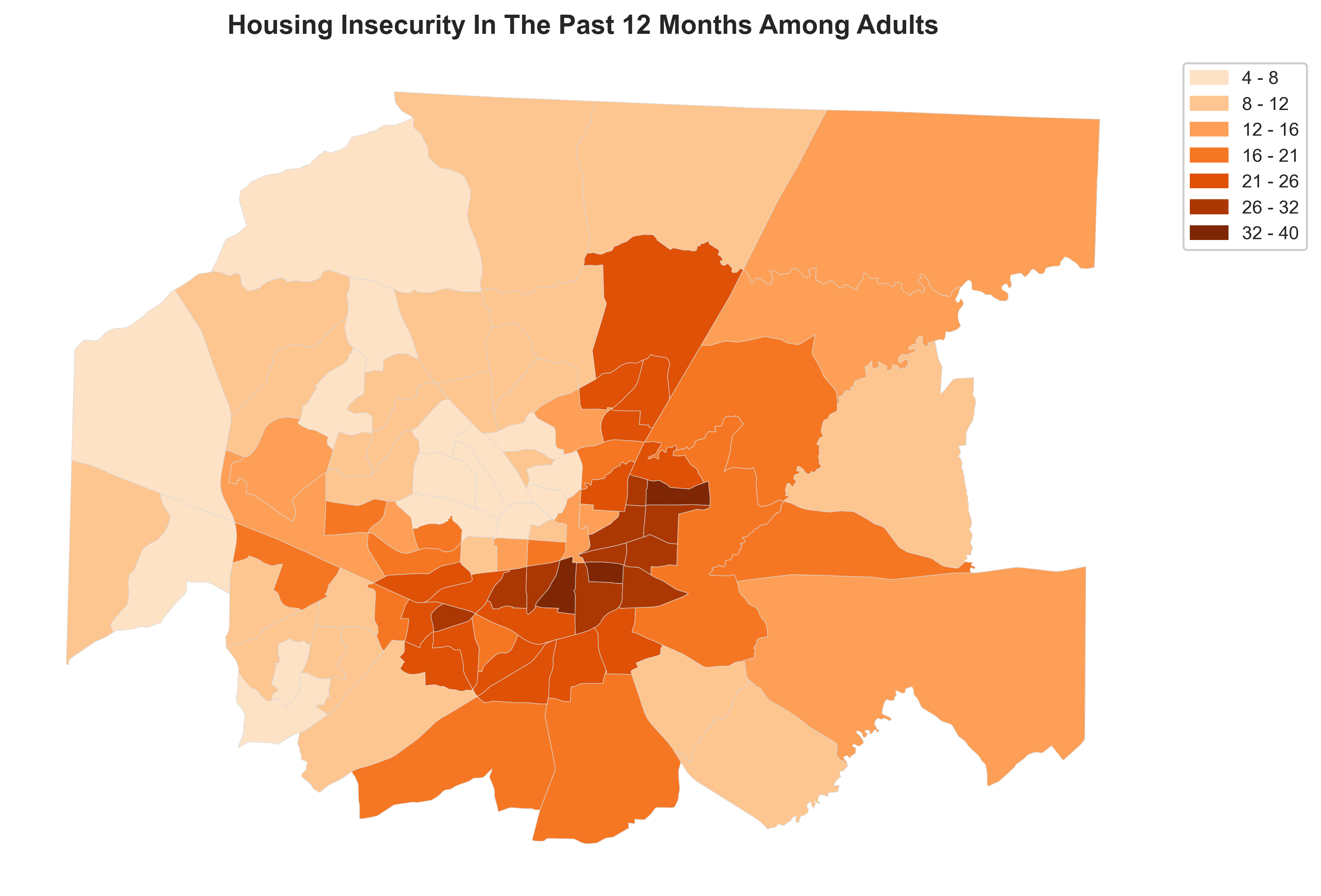} \\
\includegraphics[width=0.23\textwidth]{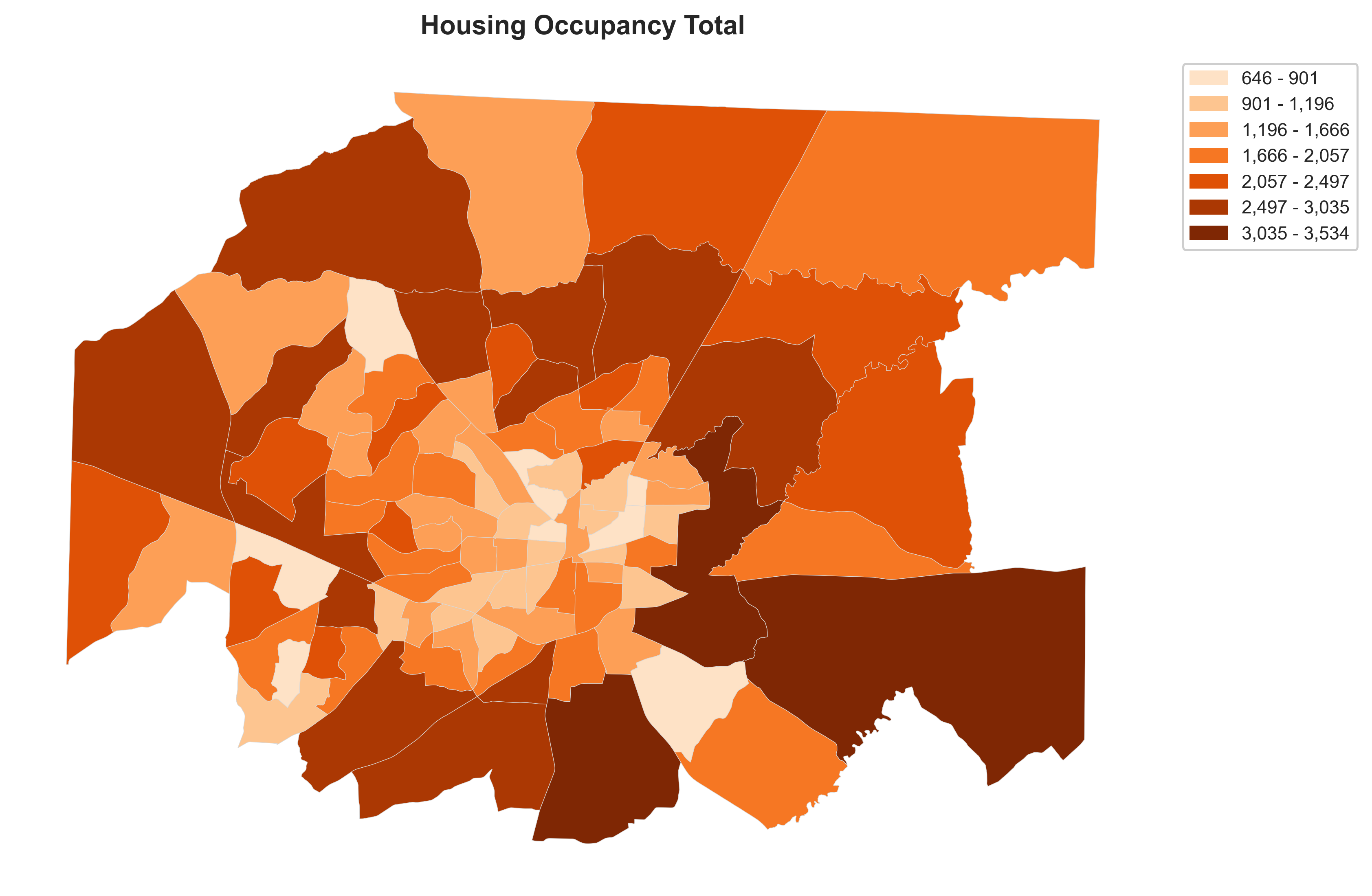} &
\includegraphics[width=0.23\textwidth]{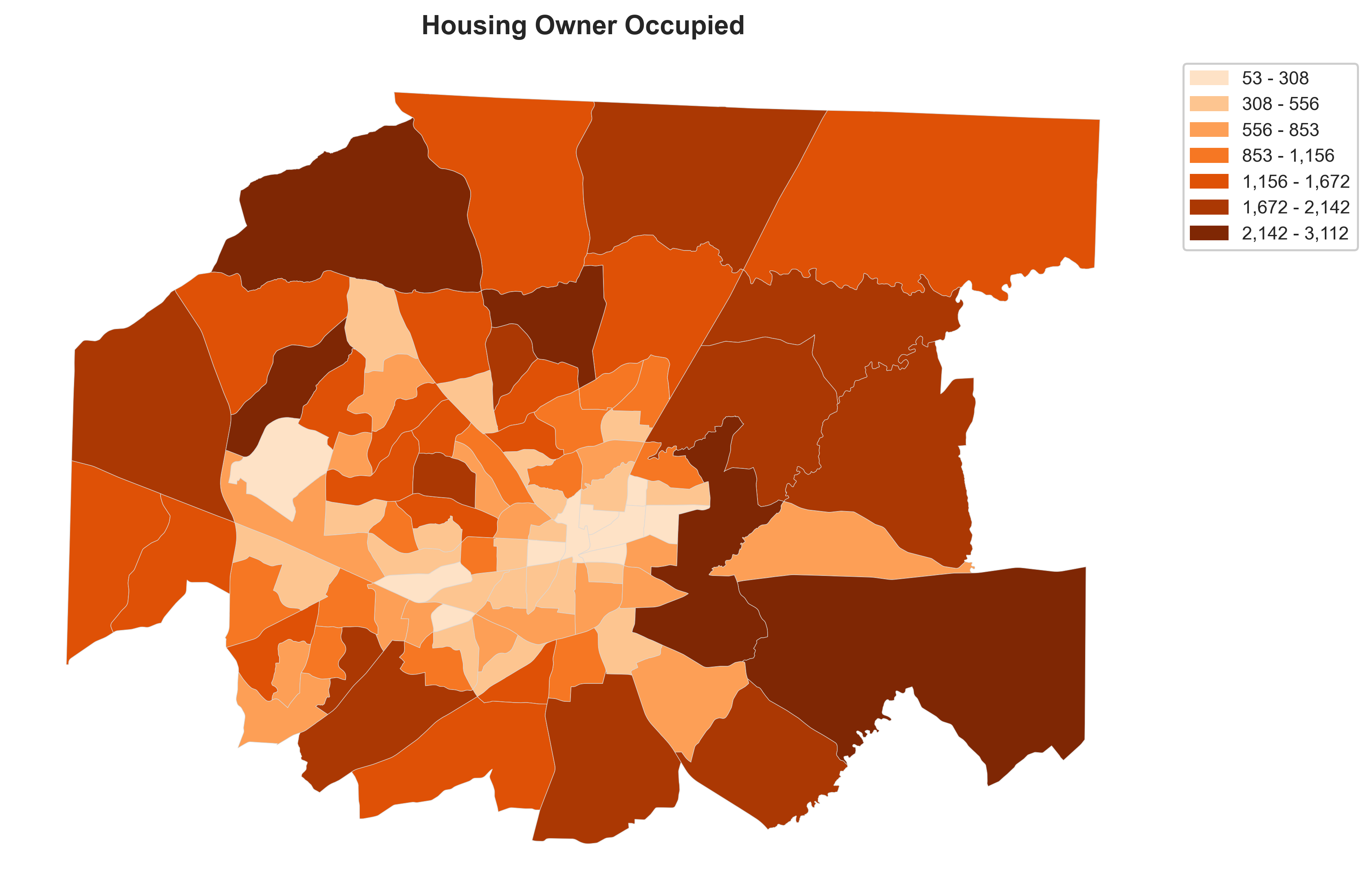} &
\includegraphics[width=0.23\textwidth]{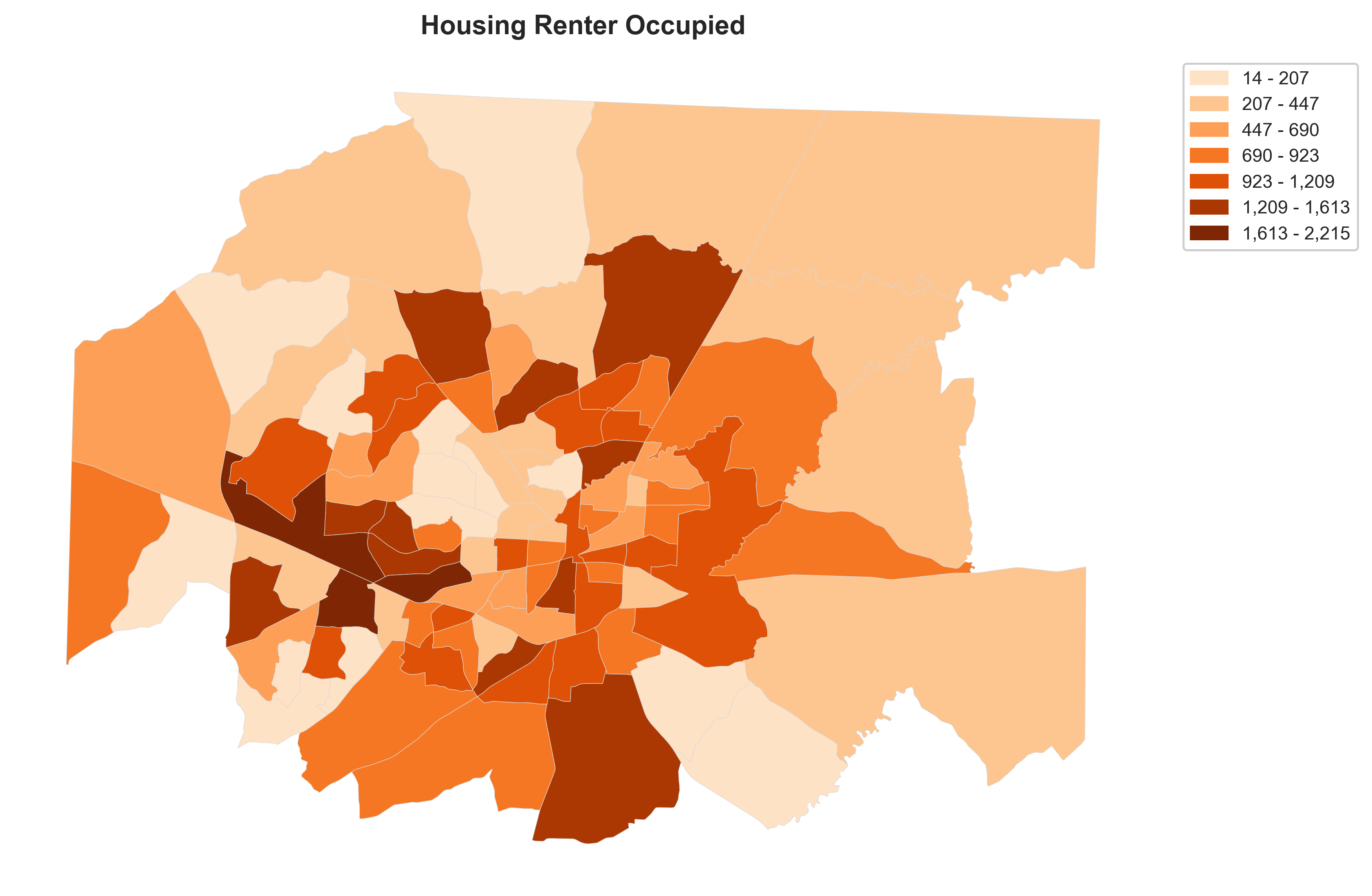} &
\includegraphics[width=0.23\textwidth]{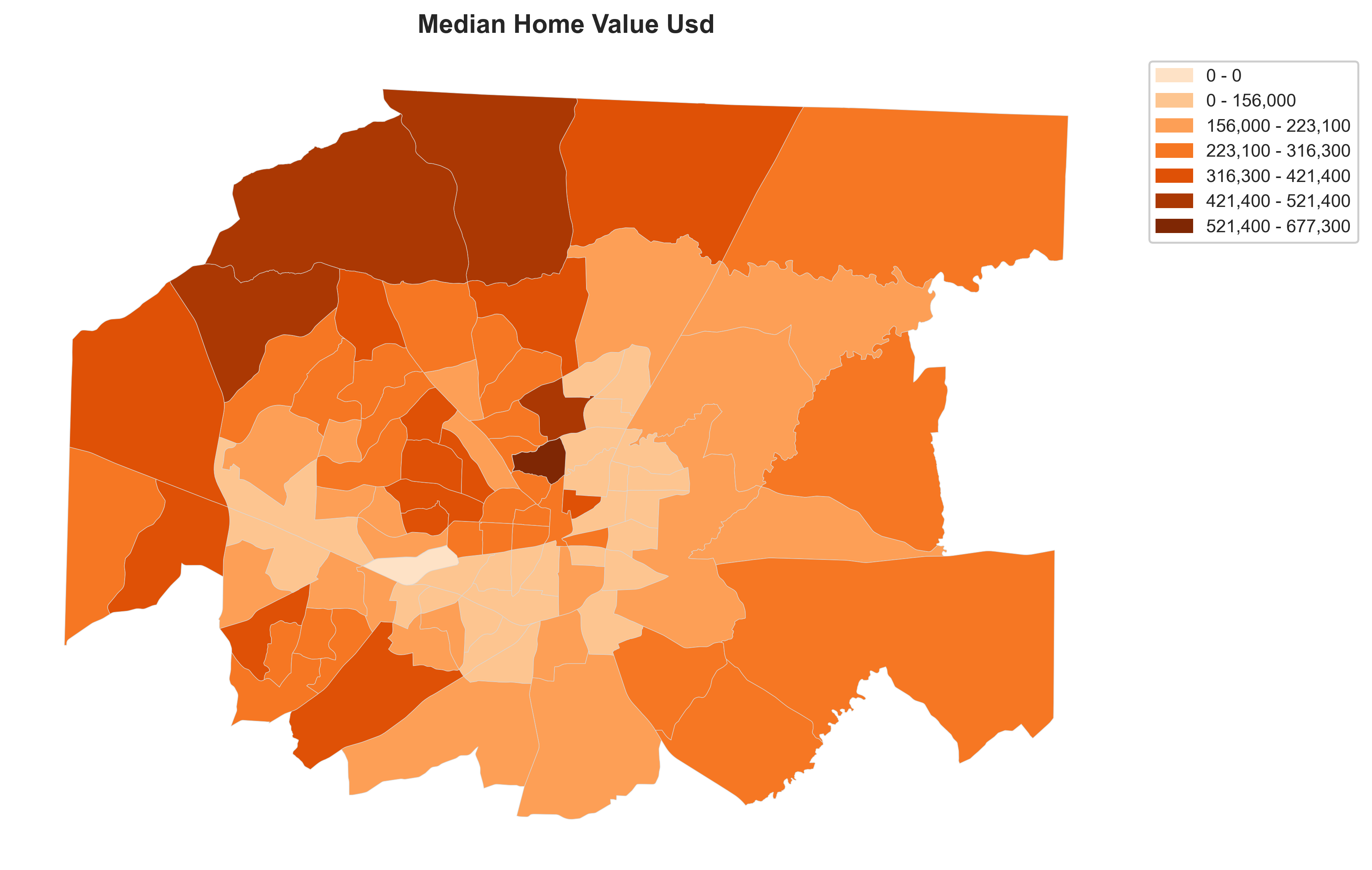} \\
\includegraphics[width=0.23\textwidth]{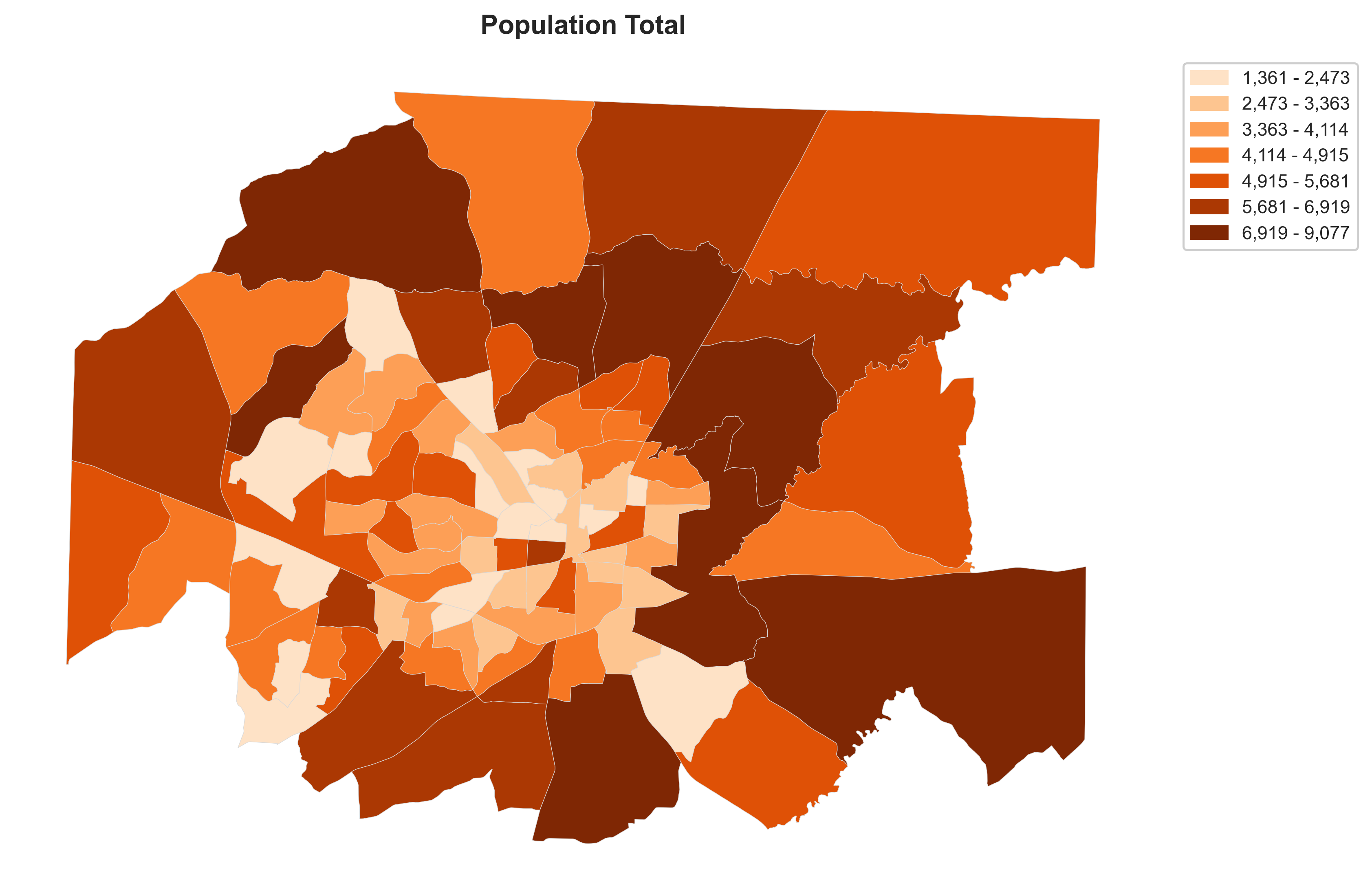} &
\includegraphics[width=0.23\textwidth]{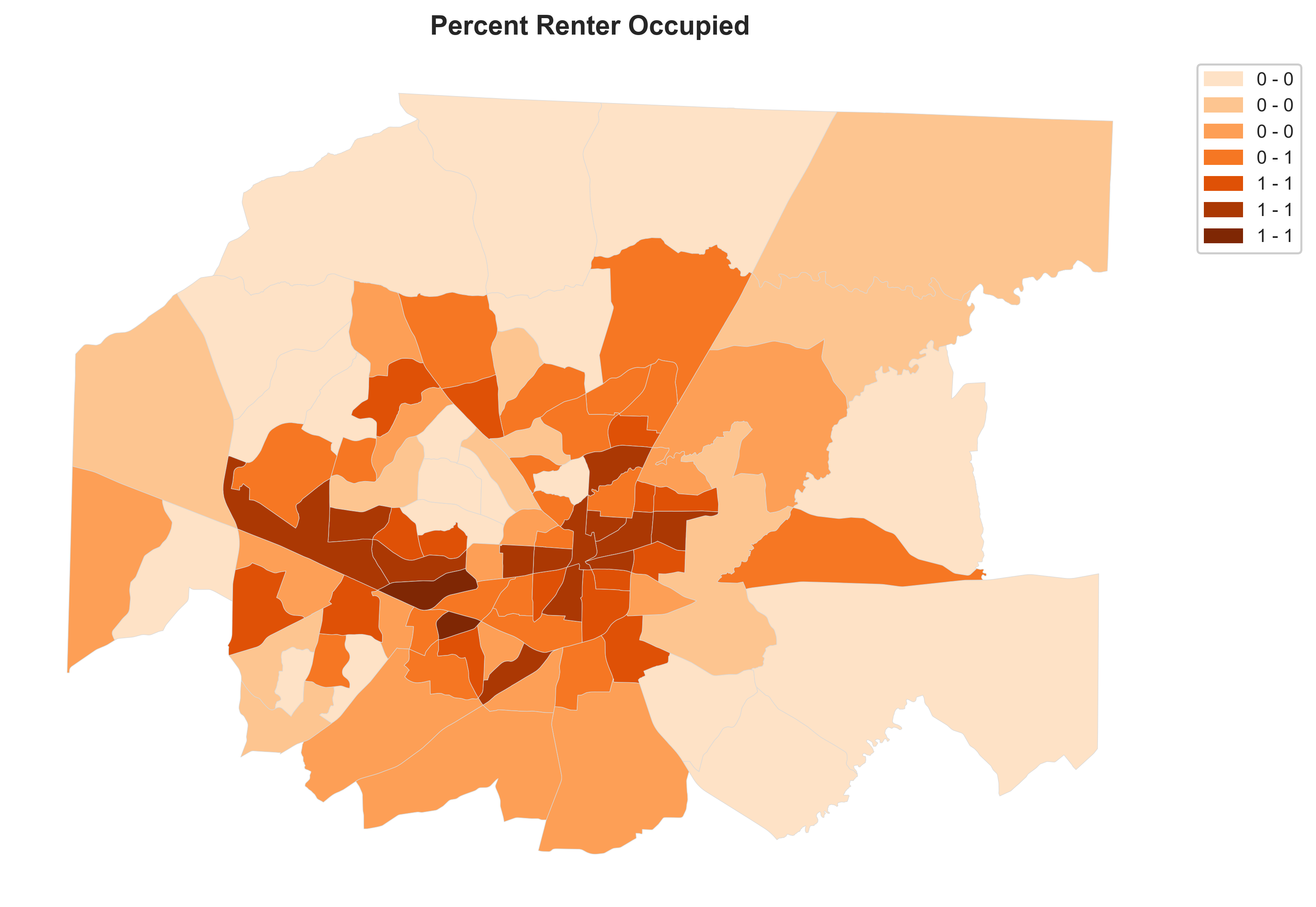} &
\includegraphics[width=0.23\textwidth]{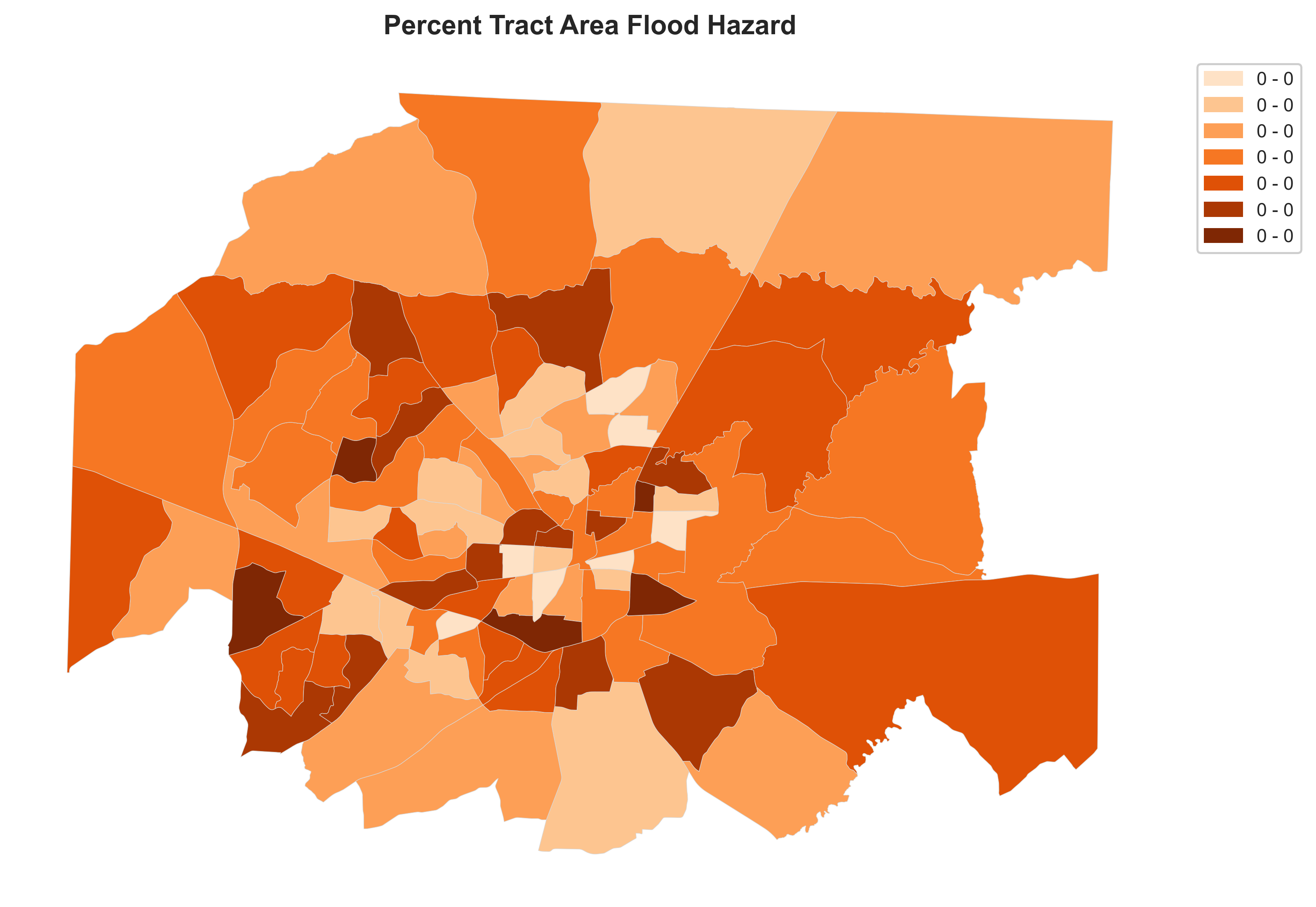} &
\includegraphics[width=0.23\textwidth]{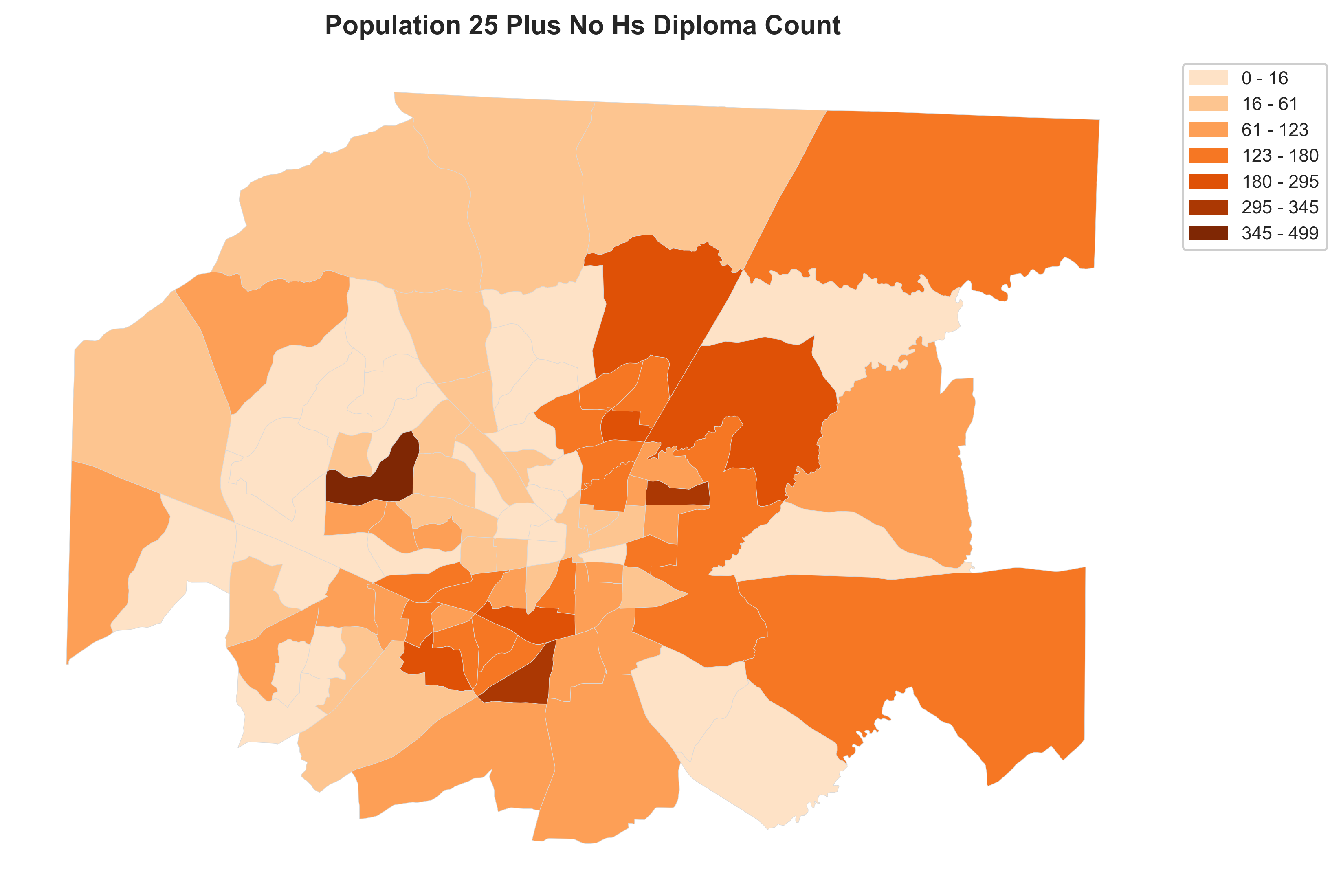} \\
\includegraphics[width=0.23\textwidth]{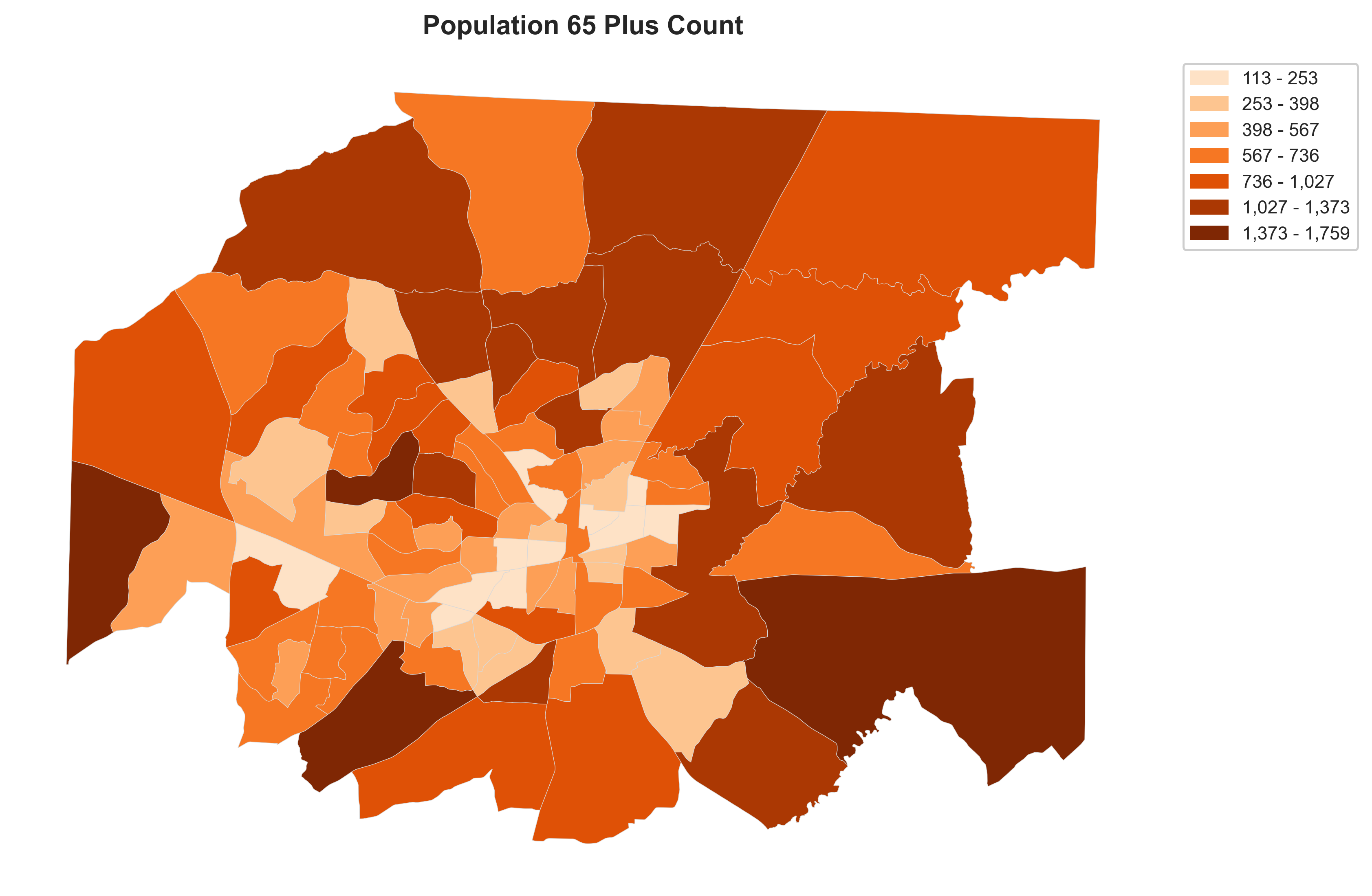} &
 &
\end{tabular}
\caption{Socioeconomic, health, housing, and environmental indicators used in the climate vulnerability assessment for Greensboro census tracts.}
\label{fig:indicator_panel}
\end{figure}

\paragraph{Overall Vulnerability Index}
The final composite vulnerability index was computed as a weighted average:
\begin{equation}
V_i = 0.6\,H_i + 0.4\,S_i.
\end{equation}
This weighting scheme emphasizes health outcomes while recognizing socioeconomic conditions as key determinants of adaptive capacity \cite{Cutter2003}. Figure~\ref{fig:indicator_panel} summarizes the primary indicators used in the assessment.

\subsection{Quality Control Procedures}

Multiple quality control measures were implemented throughout the data processing workflow. Random samples of tract-level data were cross-validated against source datasets to confirm accurate extraction and transformation. Spatial consistency was evaluated through visual inspection of choropleth maps to identify potential geographic anomalies. Statistical diagnostics, including correlation matrices and variance inflation factors, were used to assess multicollinearity among indicators. Reproducibility was ensured by documenting all processing steps within Jupyter notebooks under version control.

\section{Analytical Methods}

\subsection{Descriptive and Exploratory Analysis}

Exploratory analysis combined visual and statistical techniques to characterize the distribution of variables across Greensboro census tracts. Histograms with kernel density estimates were used to assess distributional properties, while box plots supported comparisons across poverty quartiles. Scatter plots with locally weighted scatterplot smoothing curves were employed to examine bivariate relationships.

\subsection{Correlation Analysis}

Pearson correlation coefficients were computed to quantify linear associations between variables:
\begin{equation}
r_{xy} =
\frac{\sum_{i=1}^{n}(x_i-\bar{x})(y_i-\bar{y})}
{\sqrt{\sum_{i=1}^{n}(x_i-\bar{x})^2}\sqrt{\sum_{i=1}^{n}(y_i-\bar{y})^2}},
\end{equation}
where $x_i$ and $y_i$ denote paired observations across $n$ census tracts, and $\bar{x}$ and $\bar{y}$ are their respective means. Correlation matrices were visualized using hierarchically clustered heatmaps to identify groups of variables with similar association patterns \cite{Pedregosa2011}.

\subsection{Cluster Analysis}

K-means clustering was applied to identify neighborhoods with similar vulnerability profiles using six variables representing distinct dimensions of risk: flood exposure, respiratory health burden, general health status, poverty prevalence, housing tenure instability, and aging housing stock \cite{Pedregosa2011}.

Prior to clustering, variables were standardized to z-scores:
\begin{equation}
z_{ij} = \frac{x_{ij} - \mu_j}{\sigma_j},
\end{equation}
where $\mu_j$ and $\sigma_j$ are the mean and standard deviation of variable $j$ across all tracts.

The optimal number of clusters $k$ was determined using the elbow method, which evaluates reductions in within-cluster sum of squares:
\begin{equation}
WCSS(k) = \sum_{j=1}^{k}\sum_{i \in C_j} \lVert x_i - \mu_j \rVert^2,
\end{equation}
where $C_j$ denotes cluster $j$, $x_i$ is the observation for tract $i$, and $\mu_j$ is the centroid of cluster $j$. The value of $k$ at which additional clusters yielded diminishing reductions in within-cluster variance was selected.

\subsection{Geospatial Analysis}

Spatial patterns of climate vulnerability were evaluated using complementary geospatial techniques designed to reveal both global structure and local clustering.

\paragraph{Choropleth Mapping}
Key variables and composite indices were visualized using choropleth maps classified with the Fisher--Jenks natural breaks method \cite{Jenks1967}. This classification approach minimizes within-class variance while maximizing between-class differences.

\paragraph{Spatial Autocorrelation}
Global spatial dependence was assessed using Moran’s $I$ statistic:
\begin{equation}
I =
\frac{n}{\sum_{i=1}^{n}\sum_{j=1}^{n} w_{ij}}
\frac{\sum_{i=1}^{n}\sum_{j=1}^{n} w_{ij}(x_i-\bar{x})(x_j-\bar{x})}
{\sum_{i=1}^{n}(x_i-\bar{x})^2},
\end{equation}
where $x_i$ represents the vulnerability value for tract $i$, $\bar{x}$ is the mean vulnerability across all $n$ tracts, and $w_{ij}$ denotes the spatial weight between tracts $i$ and $j$, defined using contiguity-based neighborhood relationships \cite{Anselin1995}.

\paragraph{Hot Spot Analysis}
Local Indicators of Spatial Association were computed to identify statistically significant clusters of high and low vulnerability values \cite{Anselin1995}.

\subsection{Environmental Justice Analysis}

To assess disparities in climate risk exposure, demographic and socioeconomic characteristics of high-vulnerability census tracts were compared with city-wide averages. Indicators examined included the proportion of residents identifying as racial or ethnic minorities, median household income, educational attainment, and age structure. Differences between high-vulnerability tracts and the broader city population were evaluated using t-tests for continuous variables and chi-square tests for categorical variables.

\section{Software and Computational Tools}

All analyses were conducted using Python version 3.10. Core libraries supported data manipulation, geospatial processing, statistical analysis, machine learning, visualization, and spatial statistics. The analytical workflow was containerized using Docker to support reproducibility across computing environments \cite{Merkel2014}. Source code and documentation were maintained in a public version-controlled repository and structured to align with FAIR data principles \cite{Wilkinson2016}. The workflow draws on widely used open-source components for tract-scale geospatial analysis and reproducible computation \cite{Pedregosa2011, Kluyver2016, Jordahl2020, Gillies2013}.

\section{Data Integration and Descriptive Analysis}

The final integrated dataset comprised 94 census tracts within Greensboro, North Carolina, incorporating 168 variables derived from multiple sources, including TIGER/Line geographic boundaries, 2023 American Community Survey five-year estimates for socioeconomic and housing characteristics, Centers for Disease Control and Prevention PLACES health indicators, and Federal Emergency Management Agency National Flood Hazard Layer flood exposure metrics \cite{TIGERLine2024, USCensusACS2023, CDCPLACES2024, FEMANFHL2024}. Initial preprocessing standardized variable names, resolved duplicate fields arising from dataset joins, and ensured consistency across sources. Variables used in subsequent analyses were organized into four analytical domains: health outcomes from PLACES, socioeconomic vulnerability indicators from the American Community Survey, housing and environmental conditions derived from the American Community Survey and FEMA data, and core demographic characteristics.

\begin{figure}[htbp]
\centering
\includegraphics[width=0.85\textwidth]{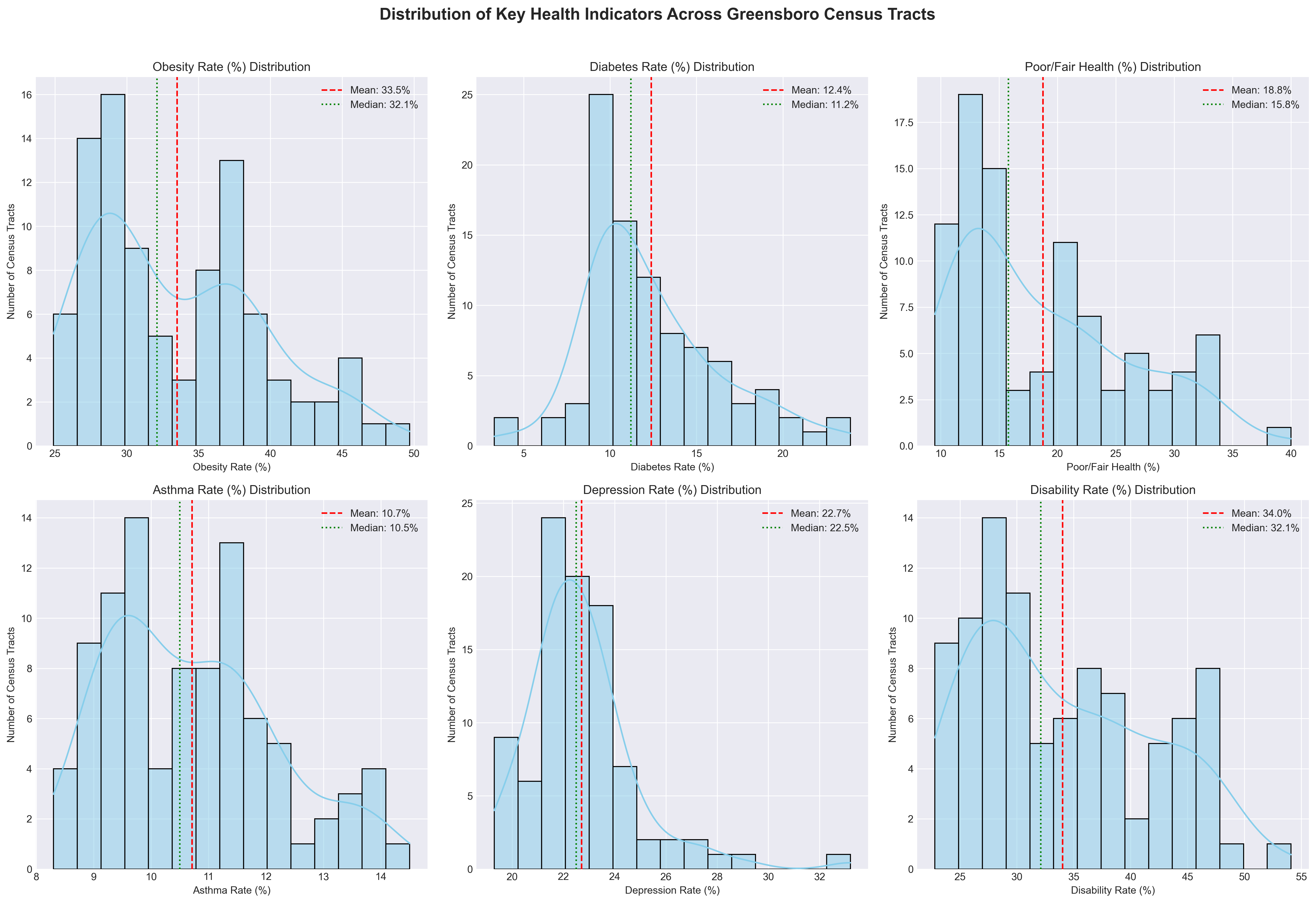}
\caption{Distribution of key health indicators across Greensboro census tracts.}
\label{fig:health_distribution}
\end{figure}

\begin{figure}[htbp]
\centering
\includegraphics[width=0.85\textwidth]{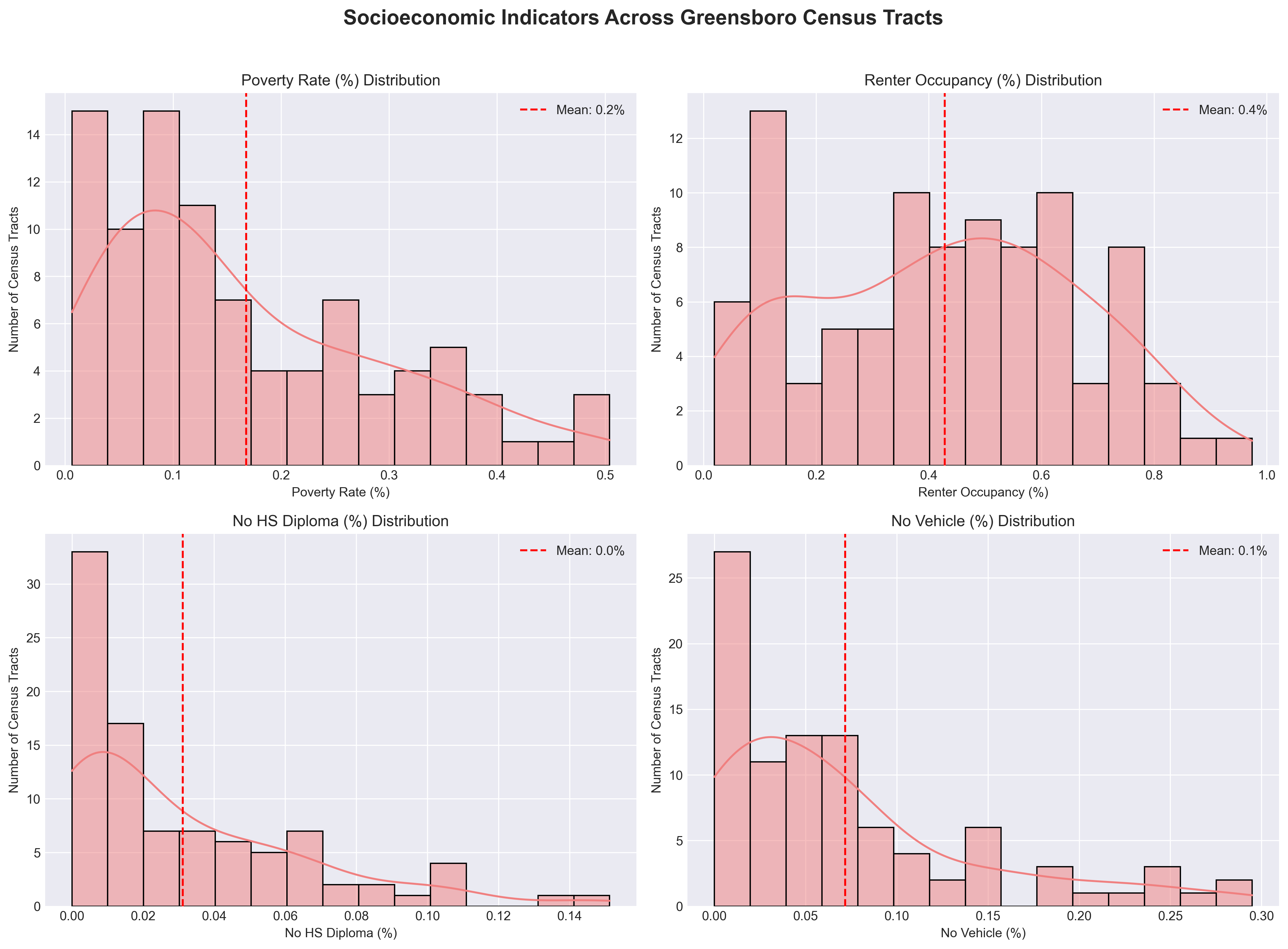}
\caption{Distribution of key socioeconomic indicators across Greensboro census tracts.}
\label{fig:socioeconomic_distribution}
\end{figure}

Descriptive analysis revealed substantial heterogeneity across census tracts. Health indicators, including adult obesity, diagnosed diabetes, and poor or fair self-rated health status, exhibited distributions skewed toward higher values (see Figure~\ref{fig:health_distribution}), reflecting a pronounced chronic disease burden in several neighborhoods. Socioeconomic indicators such as poverty prevalence, renter occupancy, and lack of a high school diploma also displayed wide ranges (see Figure~\ref{fig:socioeconomic_distribution}), confirming marked intra-urban disparities across Greensboro.

\begin{figure}[htbp]
\centering
\includegraphics[width=0.85\textwidth]{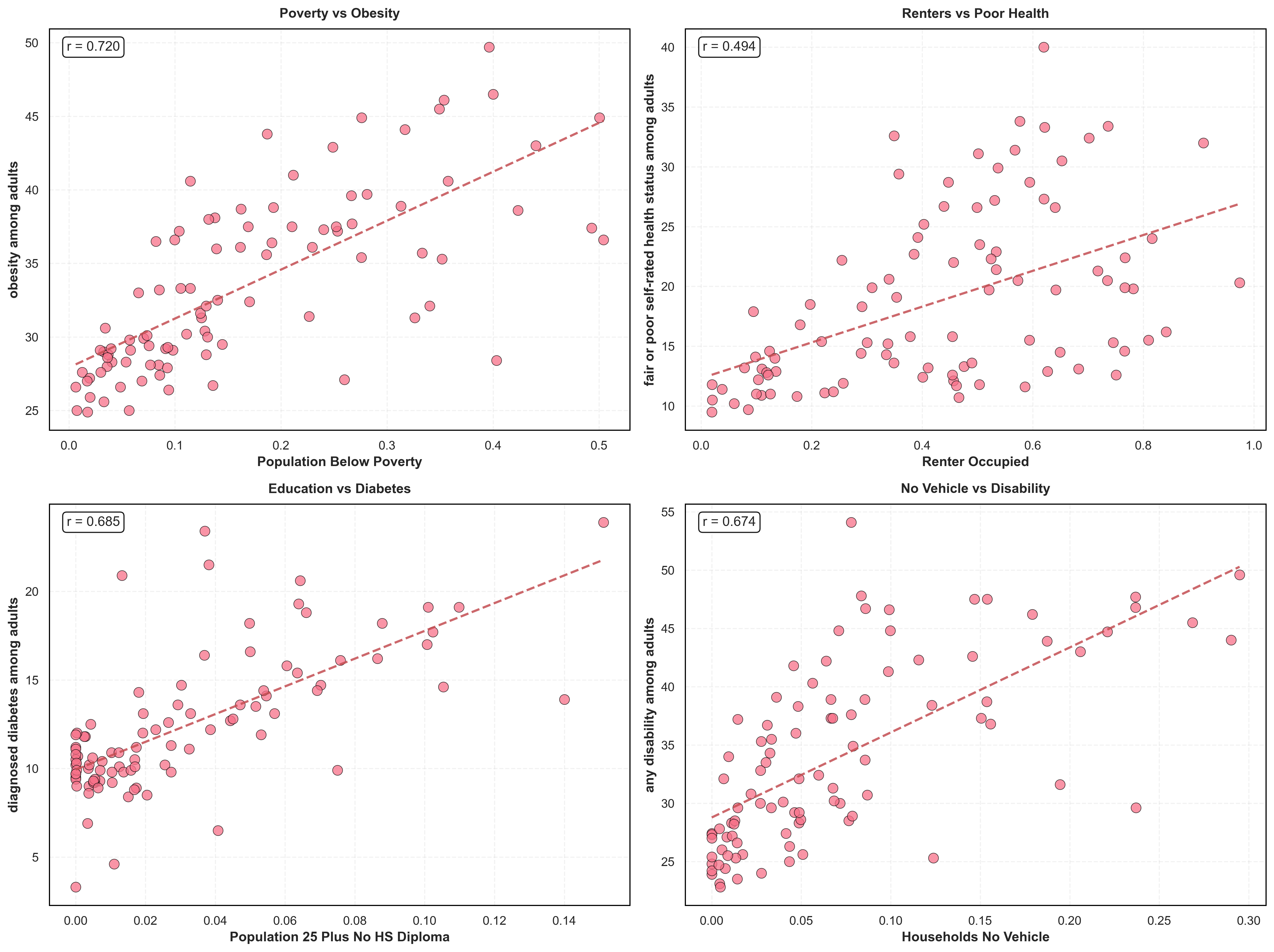}
\caption{Scatter plots illustrating relationships between health outcomes and socioeconomic factors across census tracts.}
\label{fig:health_vs_socioeconomic}
\end{figure}

Bivariate analysis confirmed strong and theoretically consistent relationships between health outcomes and socioeconomic conditions (see Figure~\ref{fig:health_vs_socioeconomic}). Poverty rate exhibited a strong positive correlation with adult obesity prevalence ($r = 0.72$) and a moderate association with diagnosed diabetes ($r = 0.44$). Renter occupancy was positively correlated with poor or fair self-rated health status ($r = 0.49$), consistent with established evidence on the social determinants of health \cite{Braveman2014}.

\begin{figure}[htbp]
\centering
\includegraphics[width=0.85\textwidth]{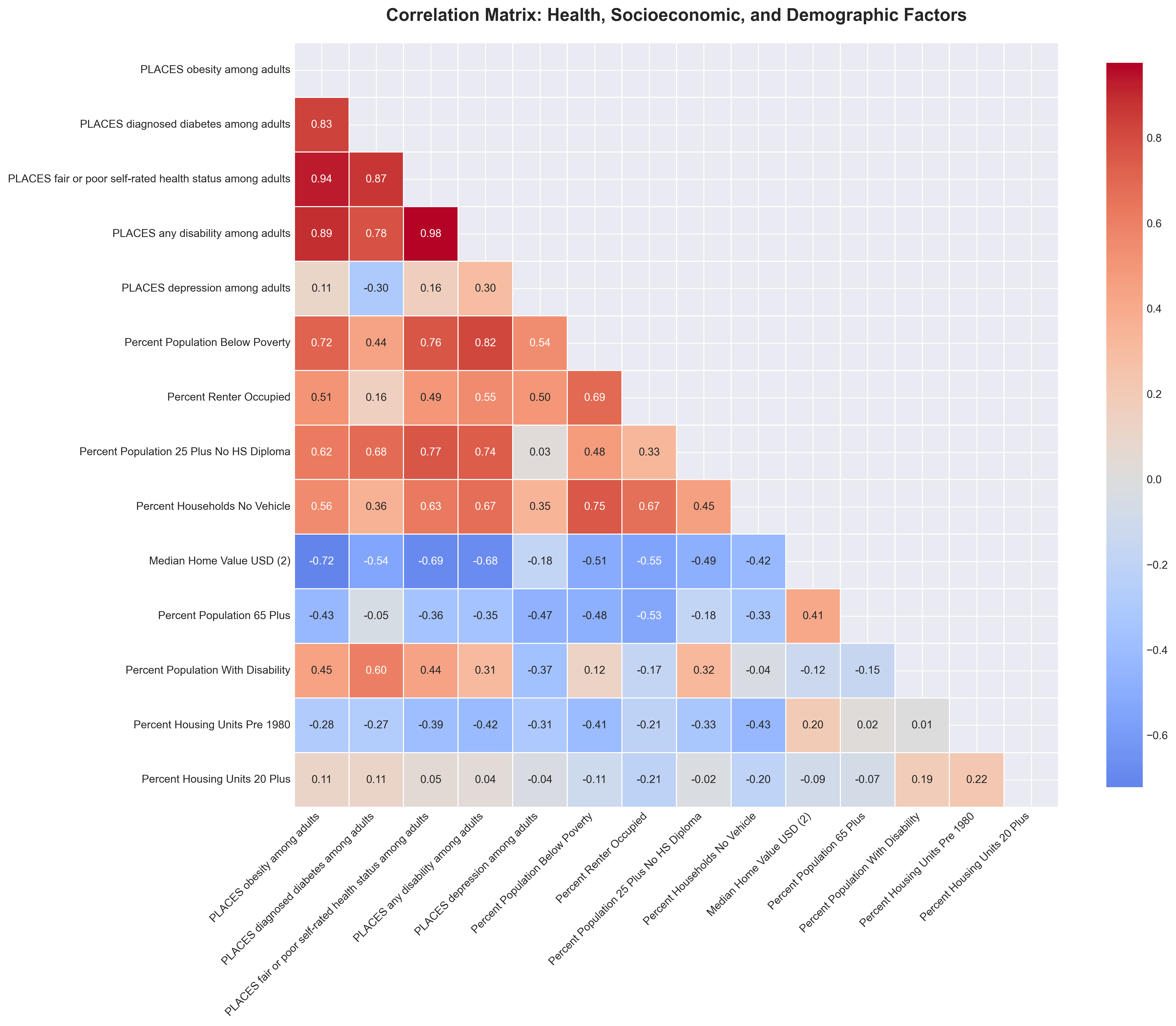}
\caption{Correlation matrix of health, socioeconomic, housing, and demographic variables.}
\label{fig:correlation_matrix}
\end{figure}

A comprehensive correlation matrix further illustrated the interconnected structure of health, socioeconomic, housing, and demographic variables (see Figure~\ref{fig:correlation_matrix}). The strongest correlations were observed among health indicators themselves, such as disability prevalence and poor health status ($r = 0.98$), as well as between poverty and multiple adverse health and housing outcomes.

\begin{figure}[htbp]
\centering
\includegraphics[width=0.85\textwidth]{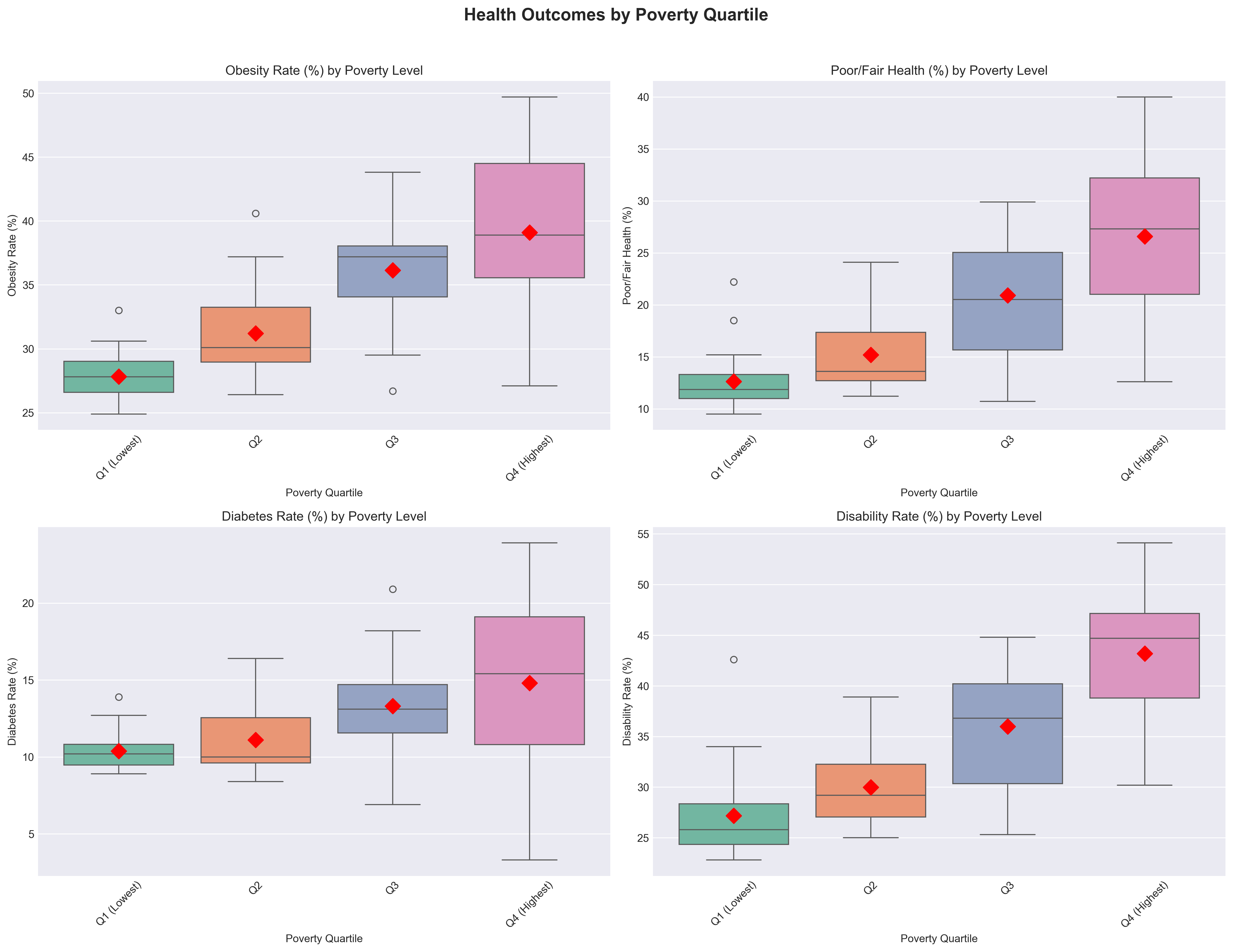}
\caption{Health outcomes stratified by poverty quartile across Greensboro census tracts.}
\label{fig:health_by_poverty}
\end{figure}

Stratifying census tracts by poverty quartile revealed a clear gradient in health outcomes (see Figure~\ref{fig:health_by_poverty}). Tracts in the highest poverty quartile consistently experienced elevated rates of obesity, diabetes, poor self-rated health, and disability relative to lower-poverty tracts.

\begin{figure}[htbp]
\centering
\includegraphics[width=0.85\textwidth]{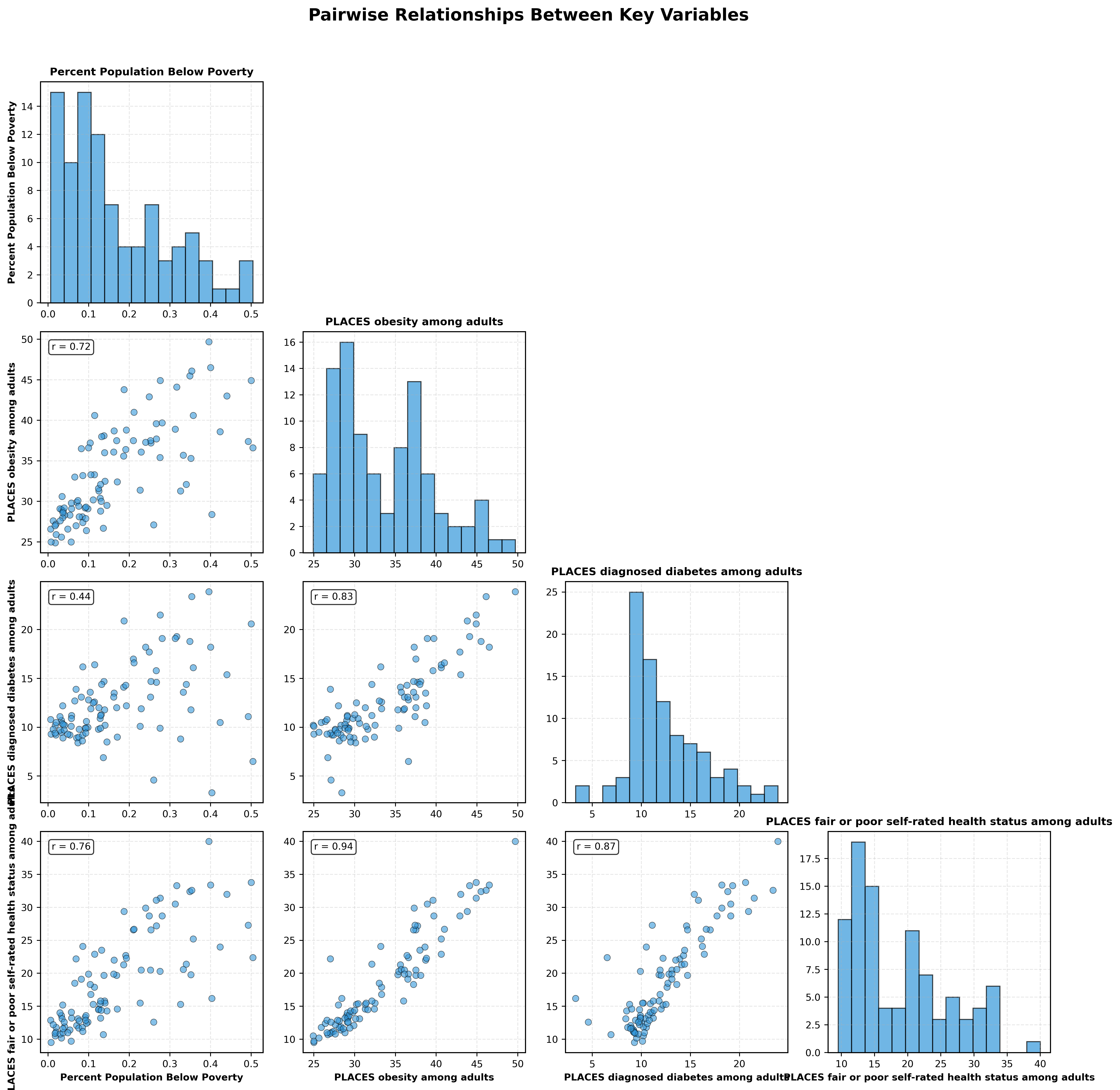}
\caption{Pairwise relationships among selected health, socioeconomic, and housing variables.}
\label{fig:pairwise_scatter}
\end{figure}

A detailed pairwise scatter matrix provided granular insight into multivariate relationships among key indicators (see Figure~\ref{fig:pairwise_scatter}), reinforcing patterns observed in the bivariate and correlation analyses.

\section{Cluster Analysis: Identifying Neighborhood Typologies}

To identify distinct neighborhood clusters facing intersecting risks, clustering was performed on a subset of 12 standardized variables representing the core analytical domains: asthma prevalence, COPD prevalence, percent flood area, percent renter-occupied, percent pre-1980 housing, poverty rate, percent with no high school diploma, percent with a disability, percent aged 65+, percent with limited English, percent households with no vehicle, and median gross rent.

\subsection{K-Means Clustering}

\begin{figure}[htbp]
\centering
\includegraphics[width=0.85\textwidth]{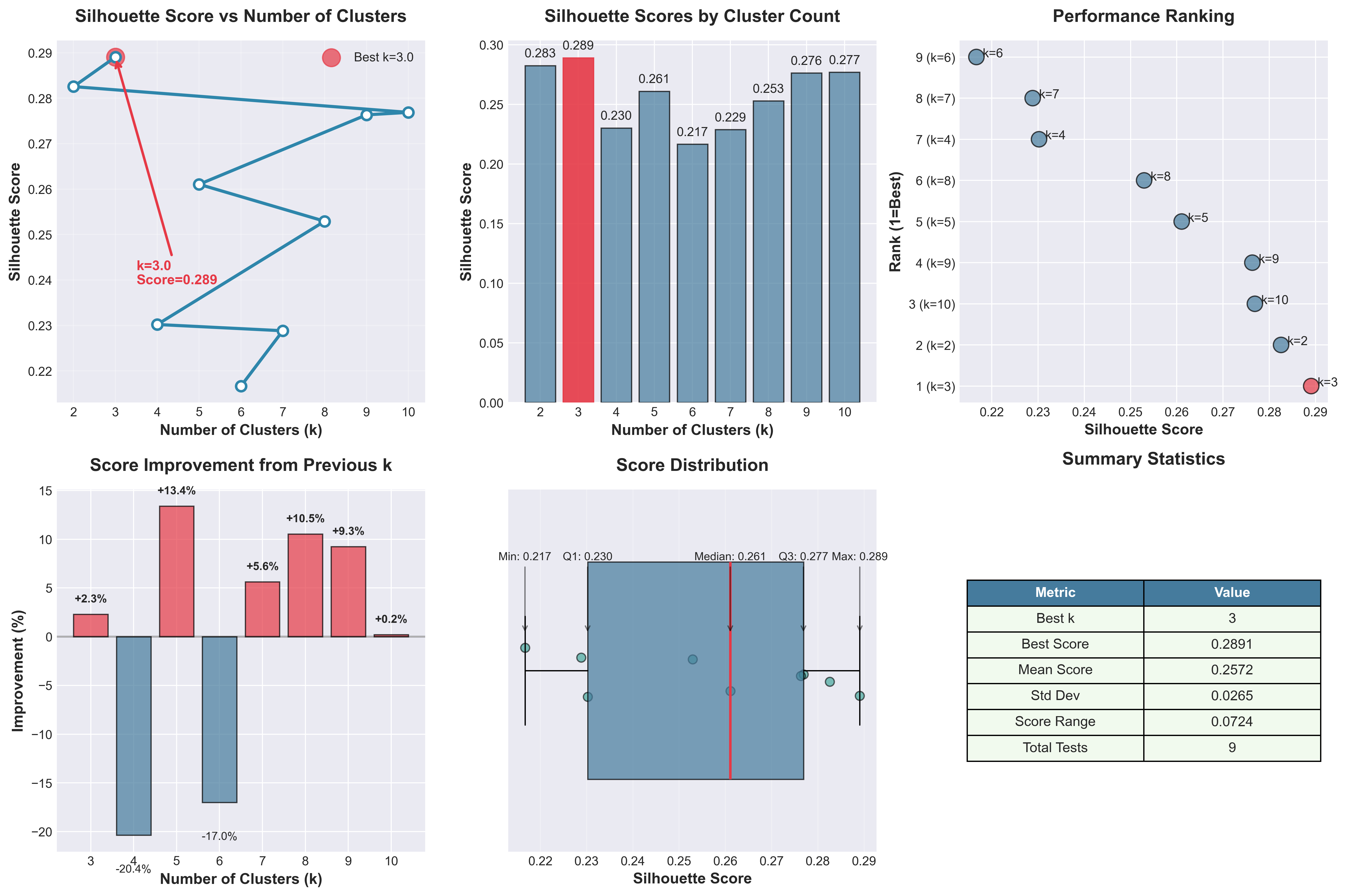}
\caption{K-means silhouette analysis for selecting the optimal number of clusters.}
\label{fig:kmeans_silhouette}
\end{figure}

A silhouette analysis was conducted to determine the optimal number of clusters ($k$) for K-means, evaluating $k$ from 2 to 10 (see Figure~\ref{fig:kmeans_silhouette}). The analysis indicated that $k=4$ provided a good balance, with a relatively high average silhouette score and coherent cluster sizes \cite{Pedregosa2011}.

\begin{figure}[htbp]
\centering
\includegraphics[width=0.85\textwidth]{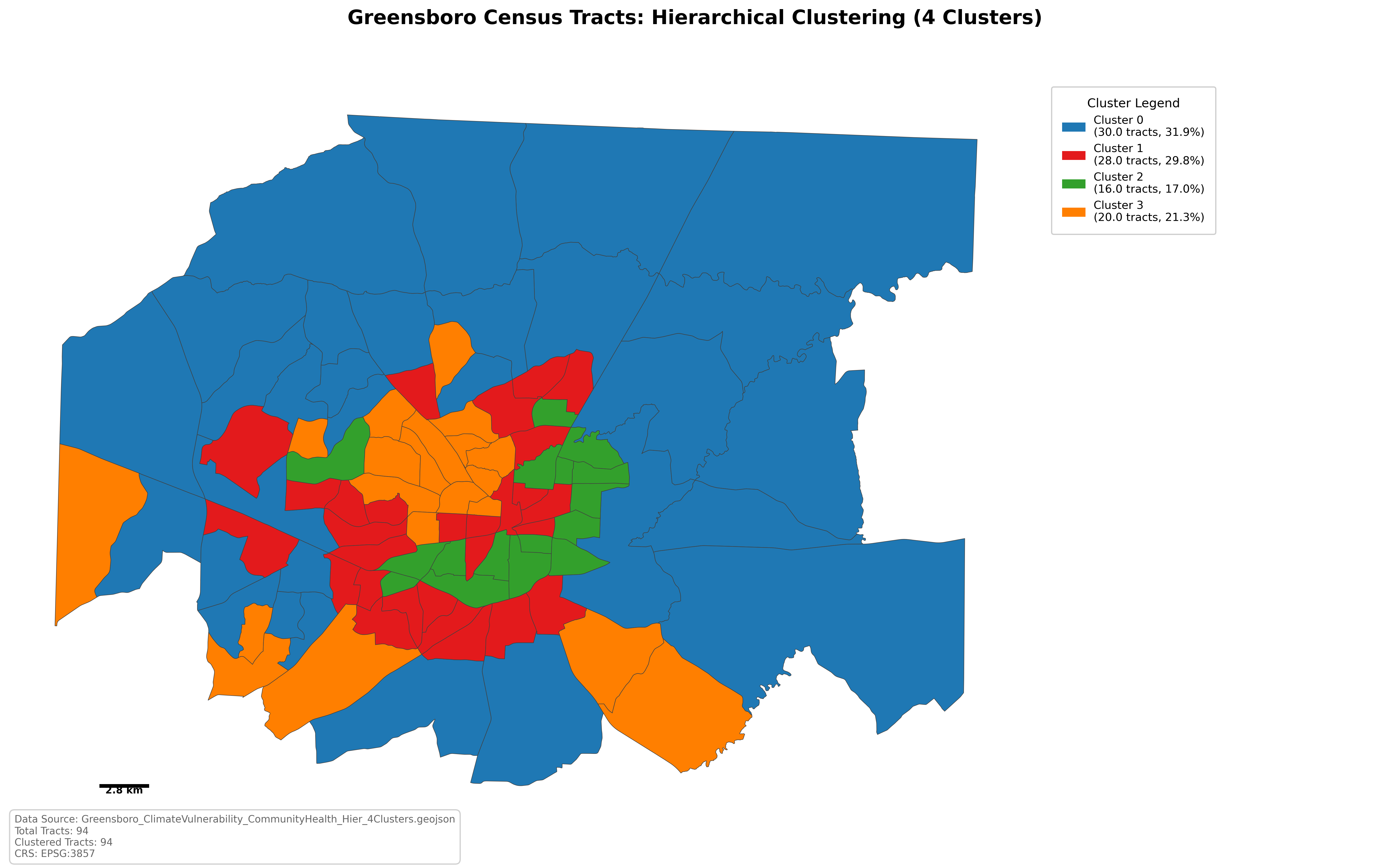}
\caption{Geographic distribution of the four K-means clusters across Greensboro census tracts.}
\label{fig:kmeans_map}
\end{figure}

\begin{figure}[htbp]
\centering
\includegraphics[width=0.75\textwidth]{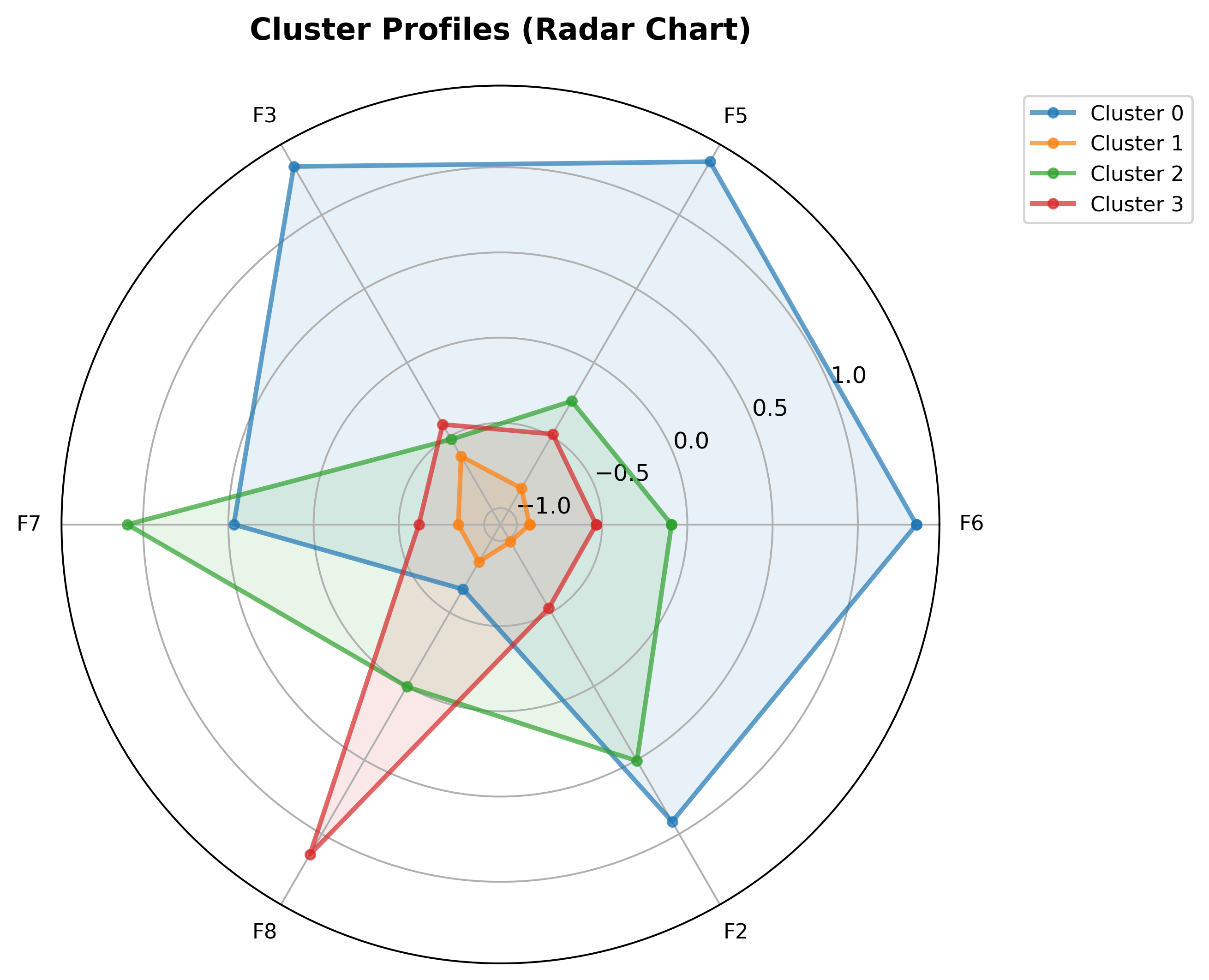}
\caption{Radar chart profile of the four K-means clusters across key indicators.}
\label{fig:kmeans_radar}
\end{figure}

The resulting four-cluster solution revealed distinct neighborhood typologies. The geographic distribution of clusters is shown in Figure~\ref{fig:kmeans_map}. A radar chart visualization shows the relative profile of each cluster across the 12 key indicators (see Figure~\ref{fig:kmeans_radar}).

\begin{figure}[htbp]
\centering
\includegraphics[width=0.85\textwidth]{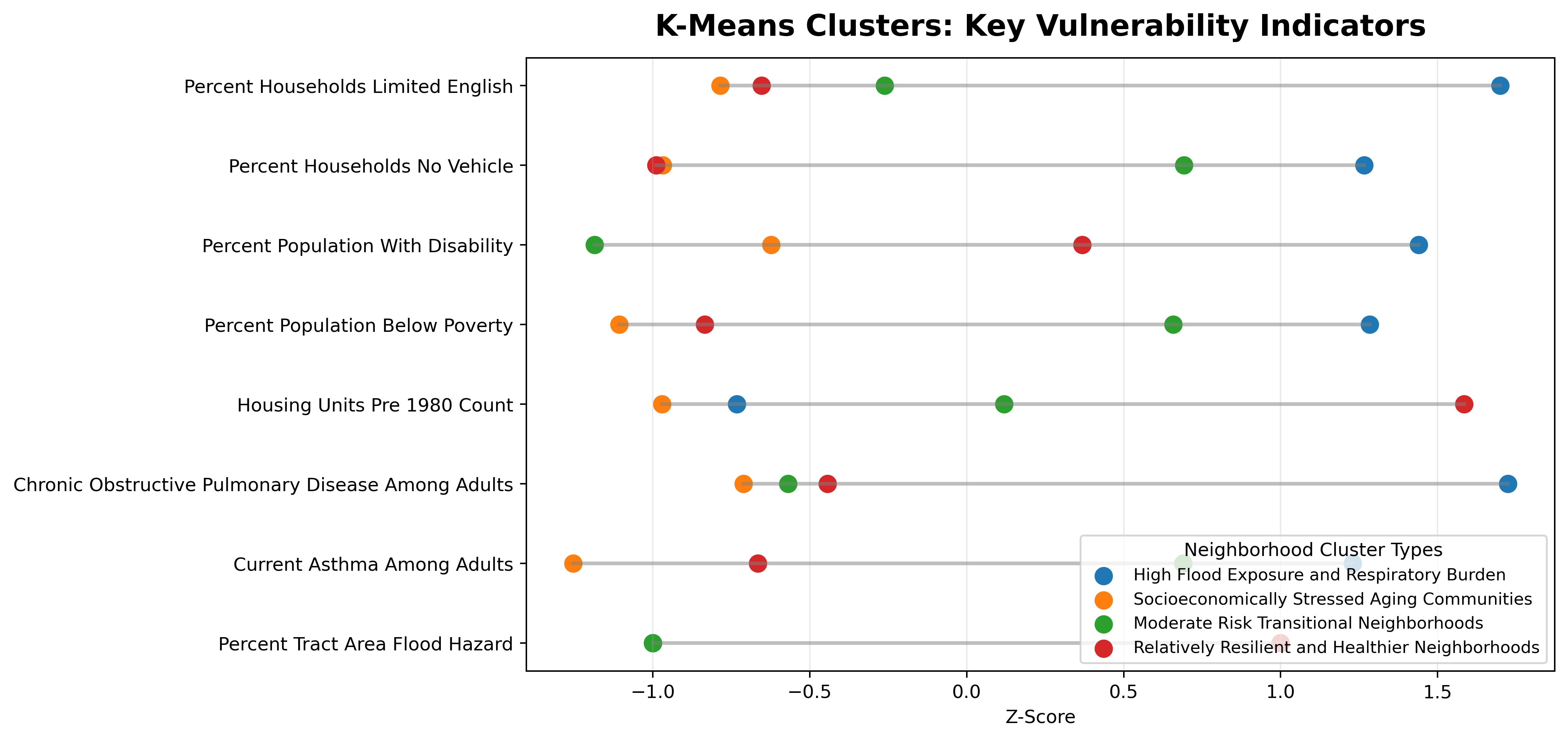}
\caption{Dumbbell chart comparing cluster extremes for selected indicators.}
\label{fig:dumbbell}
\end{figure}

Cluster means are summarized in Table~\ref{tab:cluster_means}. Key contrasts are visualized using a dumbbell chart (see Figure~\ref{fig:dumbbell}).

\begin{table}[htbp]
\centering
\caption{Mean characteristics of K-means clusters ($n=4$).}
\label{tab:cluster_means}
\begin{tabular}{lcccc}
\toprule
Characteristic &
Cluster 1 &
Cluster 2 &
Cluster 3 &
Cluster 4 \\
\midrule
Asthma prevalence (\%) & 12.1 & 9.4 & 13.5 & 10.4 \\
COPD prevalence (\%) & 8.6 & 4.8 & 9.8 & 5.8 \\
Flood area (\%) & 1.02 & 0.21 & 0.18 & 0.65 \\
Renter-occupied (\%) & 72.3 & 15.4 & 38.1 & 58.9 \\
Pre-1980 housing (\%) & 85.1 & 31.2 & 52.7 & 76.4 \\
Poverty rate (\%) & 48.2 & 8.1 & 22.3 & 35.7 \\
No high school diploma (\%) & 14.2 & 1.5 & 5.8 & 10.1 \\
Disability (\%) & 32.5 & 13.1 & 25.4 & 26.8 \\
Aged 65+ (\%) & 8.7 & 25.1 & 15.3 & 12.4 \\
Limited English (\%) & 9.8 & 1.2 & 2.1 & 6.5 \\
No vehicle (\%) & 28.3 & 4.5 & 10.2 & 19.7 \\
Median gross rent (USD) & 934 & 1,245 & 1,080 & 1,012 \\
\bottomrule
\end{tabular}
\end{table}

Cluster 1 (High Vulnerability, High Flood Risk) combines the highest flood exposure with severe socioeconomic disadvantage, older housing, and elevated respiratory health burdens. Cluster 2 (Stable, Low-Risk) is characterized by low flood risk, low poverty and health burdens, higher housing stability, and the highest median rent. Cluster 3 (High Health Burden, Lower Flood Risk) exhibits the highest asthma and COPD prevalence but relatively low flood exposure and mid-range socioeconomic indicators. Cluster 4 (High Socioeconomic Stress, Moderate Flood Risk) shows substantial socioeconomic stress and moderate flood exposure, with health burdens lower than Cluster 1.

\subsection{Hierarchical Clustering (Ward's Method)}

\begin{figure}[htbp]
\centering
\includegraphics[width=0.85\textwidth]{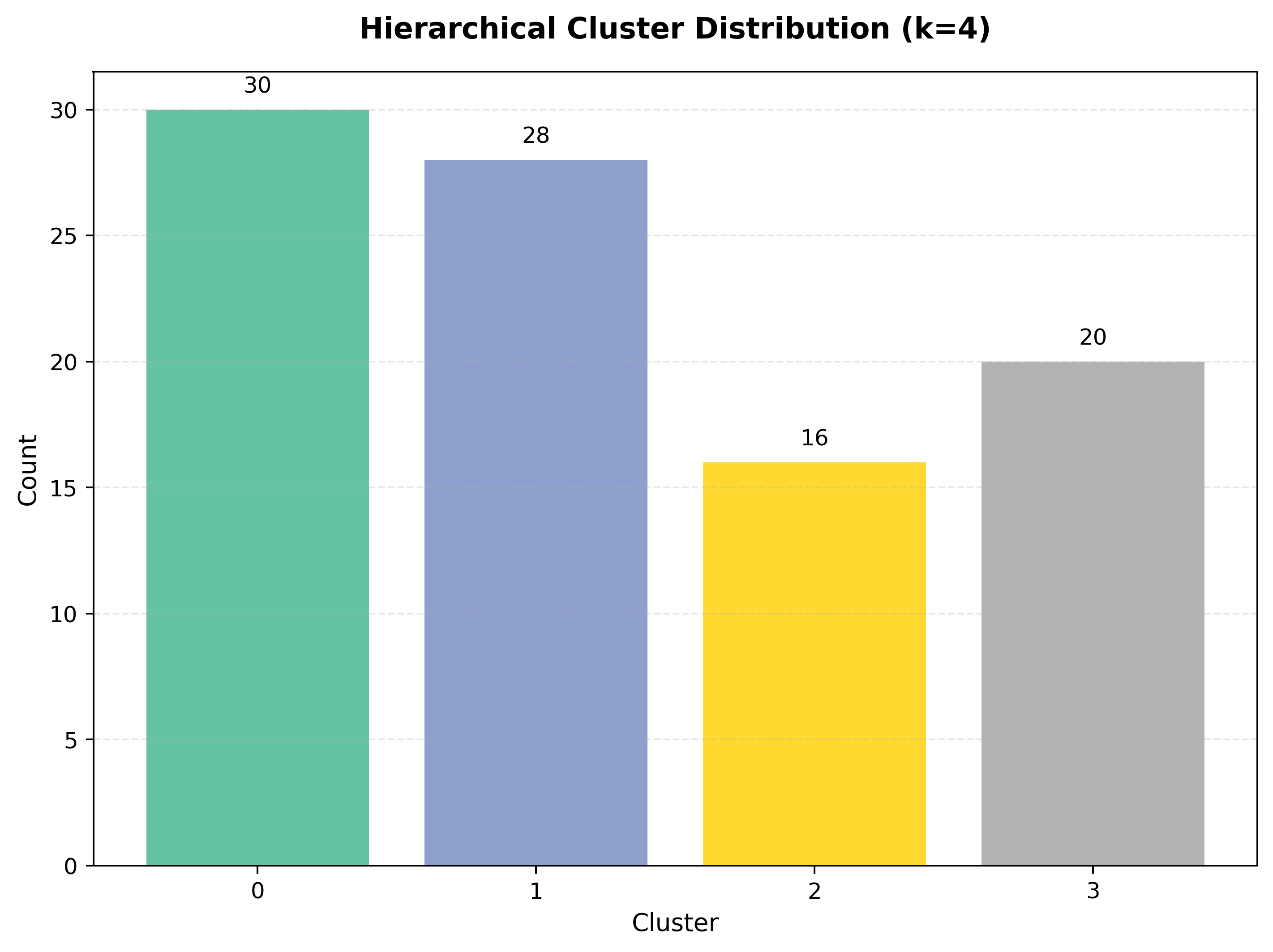}
\caption{Hierarchical cluster distribution (dendrogram cut at four clusters).}
\label{fig:hierarchical_dendrogram}
\end{figure}

\begin{figure}[htbp]
\centering
\includegraphics[width=0.85\textwidth]{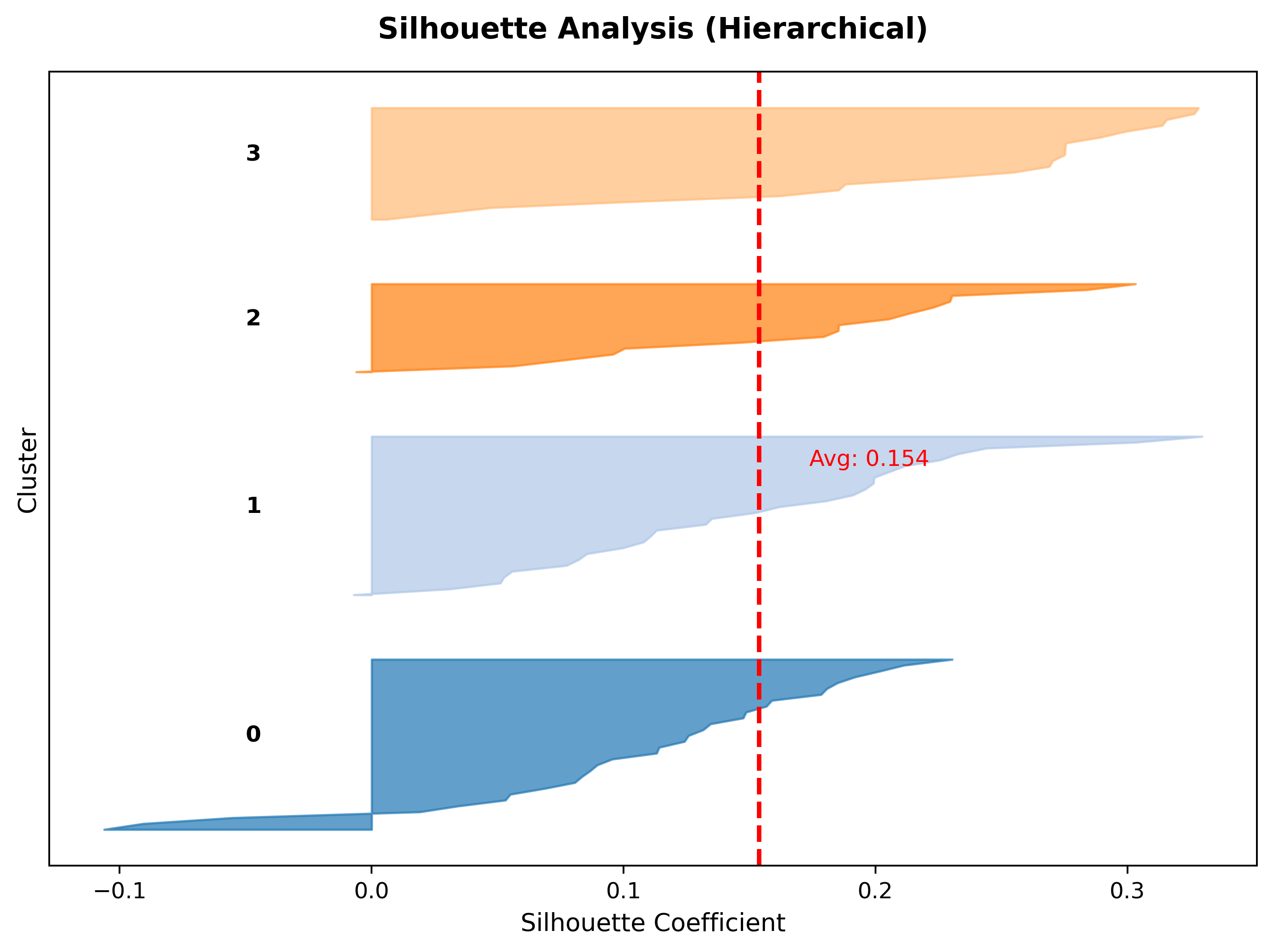}
\caption{Silhouette analysis for the four hierarchical clusters.}
\label{fig:hierarchical_silhouette}
\end{figure}

To validate robustness, agglomerative hierarchical clustering using Ward’s method was performed, specifying four clusters for comparability. The dendrogram and silhouette analysis support a four-cluster partition (see Figures~\ref{fig:hierarchical_dendrogram} and \ref{fig:hierarchical_silhouette}).

\begin{figure}[htbp]
\centering
\includegraphics[width=0.85\textwidth]{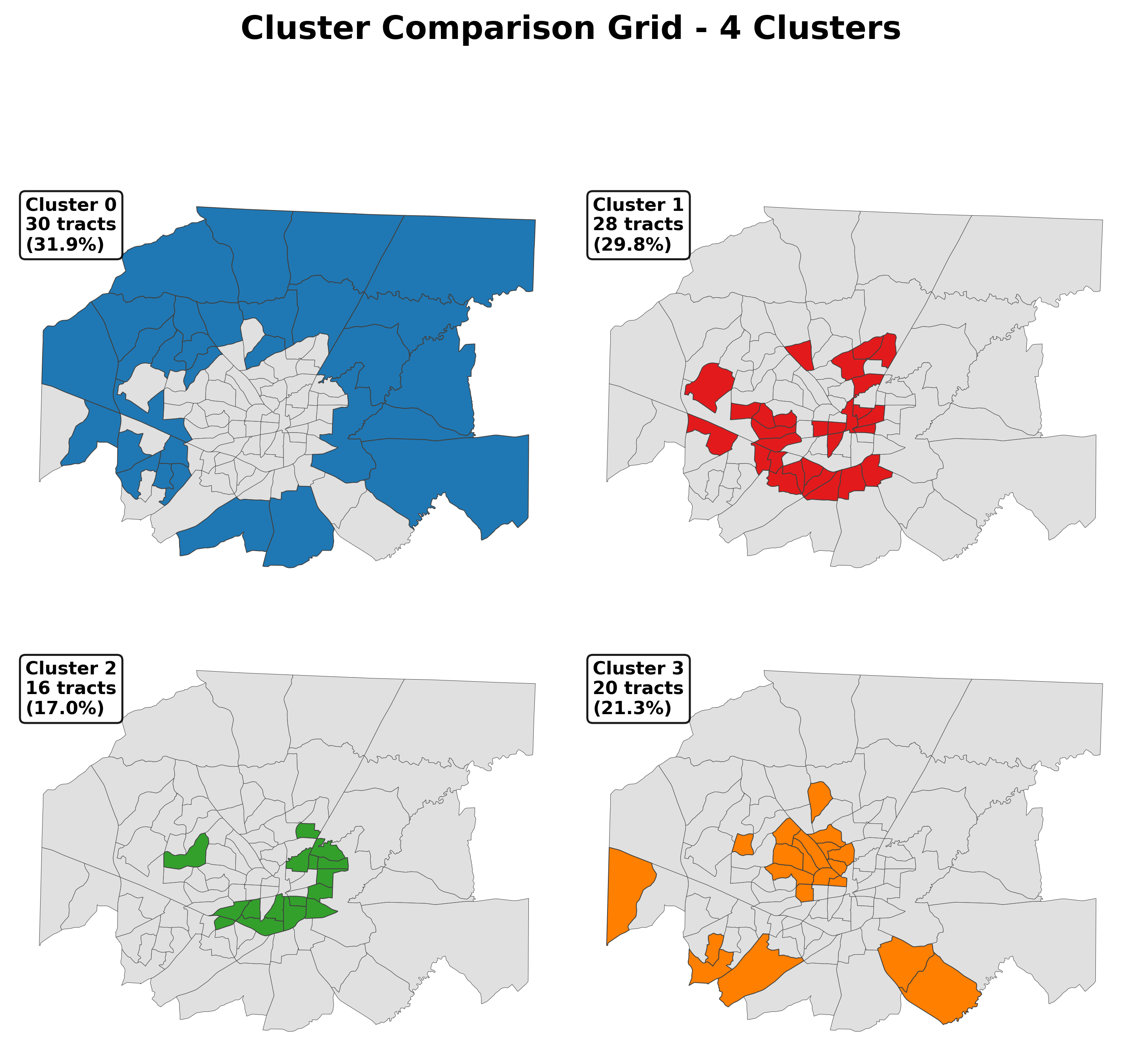}
\caption{Grid comparison of K-means and hierarchical clustering results.}
\label{fig:cluster_comparison}
\end{figure}

A grid comparison of K-means and hierarchical clustering shows strong agreement in identifying the core high-risk and low-risk areas (see Figure~\ref{fig:cluster_comparison}).

\section{Synthesis and Environmental Justice Implications}

The clustering analysis identifies neighborhoods at the intersection of climate exposure, health risk, and socioeconomic stress. The most critically vulnerable typology is Cluster 1, where flood exposure, elevated respiratory disease burden, and constrained adaptive capacity converge. This pattern is consistent with environmental justice research that links compound hazards with structural disadvantage \cite{Cutter2003, Chakraborty2019, Adepoju2021}.

Secondary vulnerable typologies include Cluster 3, which warrants targeted health and environmental interventions, and Cluster 4, where moderate flood exposure intersects with economic precarity. The Stable, Low-Risk cluster provides a benchmark illustrating how housing stability and socioeconomic advantage contribute to resilience.

\section{Discussion}

This study identified distinct neighborhood typologies in Greensboro, NC, based on the intersection of flood hazard exposure, chronic health burdens, and socioeconomic vulnerability. Our analysis reveals that climate and health risks are not uniformly distributed but are concentrated in specific, often overlapping, geographies where environmental exposure, pre-existing population sensitivity, and low adaptive capacity converge. The discussion contextualizes these findings, explores their implications for theory and practice, acknowledges limitations, and proposes pathways for intervention.

\subsection{Interpretation of Key Findings}

The identification of Cluster 1 (High Vulnerability, High Flood Risk) provides a clear, data-driven answer to the core research question. These neighborhoods epitomize compound risk \cite{Cutter2003, Kettle2016}. The high rate of pre-1980 housing suggests potential exposure to aging infrastructure less resistant to water damage. The confluence of extreme poverty, high rental tenure, and low vehicle access indicates low adaptive capacity: residents have limited resources to invest in flood-proofing, face displacement risk post-disaster due to limited ownership, and may be unable to evacuate or access post-flood services.

The strong correlation between poverty and adverse health outcomes is consistent with the social determinants of health literature \cite{Braveman2014}. This study extends that evidence by spatially layering respiratory risks with flood hazards. Post-flood environments can elevate mold, dampness, and airborne particulates, which can exacerbate respiratory conditions \cite{Quinn2017}. Therefore, residents in Cluster 1 are not only more likely to experience flooding but may also face heightened health consequences.

The contrast between Cluster 3 (High Health Burden) and Cluster 4 (High Socioeconomic Stress) highlights distinct vulnerability pathways, reinforcing that a single index can mask etiological differences. This supports place-based, typology-oriented interpretations of vulnerability \cite{Anderko2014, Reid2008}.

\subsection{Theoretical and Practical Implications}

This work aligns with a syndemic framing of climate-health interactions, where environmental hazards, chronic disease, and socioeconomic deprivation mutually reinforce one another in specific populations \cite{Singer2017}. Practically, results support targeted allocation of flood mitigation, housing protections, and health outreach to the highest-risk typologies, consistent with broader environmental justice findings \cite{Chakraborty2019, Adepoju2021}.

\subsection{Limitations and Future Research}

Key limitations include temporal misalignment across datasets, tract-level exposure aggregation that may obscure within-tract variability, unmeasured variables such as fine-scale air pollutants and healthcare access, and cross-sectional design constraints. Future work can incorporate parcel-scale exposure modeling and longitudinal analyses of neighborhood change \cite{Lewis2023}.

\subsection{Conclusion}

This study moves beyond generalized vulnerability mapping by identifying neighborhood archetypes in Greensboro facing intersecting climate and health risks. The results show that the greatest danger arises from the combined accumulation of flood exposure, chronic respiratory burden, and socioeconomic marginalization. Addressing this compounded risk requires integrated policy spanning flood mitigation, housing stability, public health, and equity-centered adaptation \cite{Cutter2003, Breakey2024}.

\nocite{*}
\bibliographystyle{unsrt}
\bibliography{references}

@article{Adepoju2021,
  author  = {Adepoju, Omolola E. and Han, Daikwon and Chae, Minji and Smith, Kendra and Gilbert, Lauren and others},
  title   = {Health disparities and climate change: The intersection of three disaster events on vulnerable communities in Houston, Texas},
  journal = {International Journal of Environmental Research and Public Health},
  year    = {2021},
  volume  = {18},
  number  = {22},
  pages   = {12135},
  doi     = {10.3390/ijerph182212135}
}

@article{Anderko2014,
  author  = {Anderko, Leslie and Davies-Cole, Janice and Strunk, Anna},
  title   = {Identifying populations at risk: Interdisciplinary environmental climate change tracking},
  journal = {Public Health Nursing},
  year    = {2014},
  volume  = {31},
  number  = {6},
  pages   = {525--533},
  doi     = {10.1111/phn.12130}
}

@article{Braveman2014,
  author  = {Braveman, Paula and Gottlieb, Laura},
  title   = {The social determinants of health: It is time to consider the causes of the causes},
  journal = {Public Health Reports},
  year    = {2014},
  volume  = {129},
  number  = {2},
  pages   = {19--31}
}

@article{Breakey2024,
  author  = {Breakey, Suellen and Hovey, Donna and Sipe, Margaret and Nicholas, Patrice K.},
  title   = {Health effects at the intersection of climate change and structural racism in the United States: A scoping review},
  journal = {Journal of Climate Change and Health},
  year    = {2024},
  volume  = {15},
  pages   = {100285},
  doi     = {10.1016/j.joclim.2023.100285}
}

@misc{CDCPLACES2024,
  author = {{Centers for Disease Control and Prevention}},
  title  = {PLACES: Local Data for Better Health, Census Tract Data 2023 Release},
  year   = {2024},
  url    = {https://www.cdc.gov/places}
}

@article{Chakraborty2019,
  author  = {Chakraborty, Jayajit and Collins, Timothy W. and Grineski, Sara E.},
  title   = {Exploring the environmental justice implications of Hurricane Harvey flooding in Greater Houston, Texas},
  journal = {American Journal of Public Health},
  year    = {2019},
  volume  = {109},
  number  = {2},
  pages   = {244--250}
}

@article{Cutter2003,
  author  = {Cutter, Susan L. and Boruff, Bryan J. and Shirley, W. Lynn},
  title   = {Social vulnerability to environmental hazards},
  journal = {Social Science Quarterly},
  year    = {2003},
  volume  = {84},
  number  = {2},
  pages   = {242--261}
}

@misc{FEMANFHL2024,
  author = {{U.S. Federal Emergency Management Agency}},
  title  = {National Flood Hazard Layer (NFHL)},
  year   = {2024},
  url    = {https://hazards.fema.gov/gis/nfhl/rest/services/public/NFHL/MapServer}
}

@article{Harlan2006,
  author  = {Harlan, Sharon L. and Brazel, Anthony J. and Prashad, Lela and Stefanov, William L. and Larsen, Larissa},
  title   = {Neighborhood microclimates and vulnerability to heat stress},
  journal = {Social Science and Medicine},
  year    = {2006},
  volume  = {63},
  number  = {11},
  pages   = {2847--2863},
  doi     = {10.1016/j.socscimed.2006.07.030}
}

@article{Harlan2012,
  author  = {Harlan, Sharon L. and Declet-Barreto, Juan H. and Stefanov, William L. and Petitti, Diana B.},
  title   = {Neighborhood effects on heat deaths: Social and environmental predictors of vulnerability in Maricopa County, Arizona},
  journal = {Environmental Health Perspectives},
  year    = {2012},
  volume  = {120},
  number  = {2},
  pages   = {197--204},
  doi     = {10.1289/ehp.1104625}
}

@article{Kettle2016,
  author  = {Kettle, Nathan P. and Dow, Kirstin},
  title   = {The role of perceived risk, uncertainty, and trust on coastal climate change adaptation planning},
  journal = {Environment and Behavior},
  year    = {2016},
  volume  = {48},
  number  = {6},
  pages   = {771--800}
}

@article{Lewis2023,
  author  = {Lewis, P. Tee and Chiu, William and Nasser, Eman and Proville, Juliette and Barone, Anthony and others},
  title   = {Characterizing vulnerabilities to climate change across the United States},
  journal = {Environment International},
  year    = {2023},
  volume  = {171},
  pages   = {107726},
  doi     = {10.1016/j.envint.2022.107726}
}

@article{May2023,
  author  = {May, Erin and Du, Pengfei and Martine, Victoria},
  title   = {Environmental justice: A case study into the heat vulnerable neighborhoods of Philadelphia},
  journal = {The Commons},
  year    = {2023},
  volume  = {2},
  pages   = {1--15}
}

@article{Pedregosa2011,
  author  = {Pedregosa, Fabian and Varoquaux, Gael and Gramfort, Alexandre and Michel, Vincent and Thirion, Bertrand and Grisel, Olivier and others},
  title   = {Scikit-learn: Machine learning in Python},
  journal = {Journal of Machine Learning Research},
  year    = {2011},
  volume  = {12},
  pages   = {2825--2830}
}

@article{Quinn2017,
  author  = {Quinn, Alexander and Tamerius, James D. and Perzanowski, Matthew and Jacobson, Jeffrey S. and Goldstein, Ilan and Acosta, Laura and Shaman, Jeffrey},
  title   = {Predicting indoor heat exposure risk during extreme heat events},
  journal = {Science of the Total Environment},
  year    = {2017},
  volume  = {575},
  pages   = {1--10}
}

@article{Reid2008,
  author  = {Reid, Colleen E. and O'Neill, Marie S. and Gronlund, Charlotte J. and Brines, Shannon J. and Brown, Daniel G. and others},
  title   = {Mapping community determinants of heat vulnerability},
  journal = {Environmental Health Perspectives},
  year    = {2008},
  volume  = {117},
  number  = {11},
  pages   = {1730--1736},
  doi     = {10.1289/ehp.0900683}
}

@article{Singer2017,
  author  = {Singer, Merrill and Bulled, Nicholas and Ostrach, Bayla and Mendenhall, Emily},
  title   = {Syndemics and the biosocial conception of health},
  journal = {The Lancet},
  year    = {2017},
  volume  = {389},
  number  = {10072},
  pages   = {941--950}
}

@misc{TIGERLine2024,
  author = {{U.S. Census Bureau}},
  title  = {TIGER/Line Shapefiles},
  year   = {2024},
  url    = {https://www.census.gov/geographies/mapping-files/time-series/geo/tiger-line-file.html}
}

@misc{USCensusACS2023,
  author = {{U.S. Census Bureau}},
  title  = {American Community Survey 5-Year Estimates (2019--2023)},
  year   = {2023},
  url    = {https://www.census.gov/data/developers/data-sets/acs-5year.html}
}

@article{Voelkel2018,
  author  = {Voelkel, Jackson and Hellman, Dana and Sakuma, Ryuichi and Shandas, Vivek},
  title   = {Assessing vulnerability to urban heat: A study of disproportionate heat exposure and access to refuge by socio-demographic status in Portland, Oregon},
  journal = {International Journal of Environmental Research and Public Health},
  year    = {2018},
  volume  = {15},
  number  = {4},
  pages   = {640},
  doi     = {10.3390/ijerph15040640}
}

@article{Yu2021,
  author  = {Yu, Jessica and Castellani, Katherine and Forysinski, Krista and Gustafson, Paul and Lu, James and others},
  title   = {Geospatial indicators of exposure, sensitivity, and adaptive capacity to assess neighbourhood variation in vulnerability to climate change-related health hazards},
  journal = {Environmental Health},
  year    = {2021},
  volume  = {20},
  number  = {1},
  pages   = {31},
  doi     = {10.1186/s12940-021-00715-0}
}

@article{Gaither2021,
  author  = {Gaither, Cassandra J. and Zarnoch, Stanley J. and Abt, Karen L.},
  title   = {Climate change vulnerability and social vulnerability in the Southeast United States},
  journal = {Environmental Research Letters},
  year    = {2021},
  volume  = {16},
  number  = {3},
  pages   = {034011},
  doi     = {10.1088/1748-9326/abe467}
}

@misc{USCensusGEOID2020,
  author = {{U.S. Census Bureau}},
  title  = {Understanding Geographic Identifiers (GEOIDs)},
  year   = {2020},
  url    = {https://www.census.gov/programs-surveys/geography/guidance/geo-identifiers.html}
}

@article{Miranda2011,
  author  = {Miranda, Marie Lynn and Edwards, Sharon E. and Keating, Martha H. and Paul, Christopher J.},
  title   = {Making the environmental justice grade: The relative burden of air pollution exposure in the United States},
  journal = {International Journal of Environmental Research and Public Health},
  year    = {2011},
  volume  = {8},
  number  = {6},
  pages   = {1755--1771},
  doi     = {10.3390/ijerph8061755}
}

@article{Spielman2014,
  author  = {Spielman, Seth E. and Folch, David and Nagle, Nicholas},
  title   = {Patterns and causes of uncertainty in the American Community Survey},
  journal = {Applied Geography},
  year    = {2014},
  volume  = {46},
  pages   = {147--157},
  doi     = {10.1016/j.apgeog.2013.11.002}
}

@article{Zhang2021,
  author  = {Zhang, Xin and Holt, James B. and Lu, Hua and Wheaton, Anne G. and Ford, Earl S. and Greenlund, Kurt J. and Croft, Janet B.},
  title   = {Multilevel regression and poststratification for small-area estimation of population health outcomes: A case study of chronic obstructive pulmonary disease prevalence using the Behavioral Risk Factor Surveillance System},
  journal = {American Journal of Epidemiology},
  year    = {2021},
  volume  = {190},
  number  = {8},
  pages   = {1613--1619},
  doi     = {10.1093/aje/kwab043}
}

@book{Tukey1977,
  author    = {Tukey, John W.},
  title     = {Exploratory Data Analysis},
  publisher = {Addison-Wesley},
  year      = {1977}
}

@article{Osborne2010,
  author  = {Osborne, Jason W.},
  title   = {Improving your data transformations: Applying the Box-Cox transformation},
  journal = {Practical Assessment, Research, and Evaluation},
  year    = {2010},
  volume  = {15},
  number  = {1},
  pages   = {12},
  doi     = {10.7275/qbpc-gk17}
}

@article{Jenks1967,
  author  = {Jenks, George F.},
  title   = {The data model concept in statistical mapping},
  journal = {International Yearbook of Cartography},
  year    = {1967},
  volume  = {7},
  pages   = {186--190}
}

@article{Anselin1995,
  author  = {Anselin, Luc},
  title   = {Local indicators of spatial association (LISA)},
  journal = {Geographical Analysis},
  year    = {1995},
  volume  = {27},
  number  = {2},
  pages   = {93--115},
  doi     = {10.1111/j.1538-4632.1995.tb00338.x}
}

@misc{Gillies2013,
  author = {Gillies, Sean and others},
  title  = {Rasterio: Geospatial raster I/O for {Python}},
  year   = {2013},
  url    = {https://rasterio.readthedocs.io/}
}

@inproceedings{Kluyver2016,
  author    = {Kluyver, Thomas and Ragan-Kelley, Benjamin and P{\'e}rez, Fernando and Granger, Brian and Bussonnier, Matthias and others},
  title     = {Jupyter Notebooks: A publishing format for reproducible computational workflows},
  booktitle = {Positioning and Power in Academic Publishing: Players, Agents and Agendas},
  year      = {2016},
  pages     = {87--90},
  publisher = {IOS Press},
  doi       = {10.3233/978-1-61499-649-1-87}
}

@article{Merkel2014,
  author  = {Merkel, Dirk},
  title   = {Docker: Lightweight Linux containers for consistent development and deployment},
  journal = {Linux Journal},
  year    = {2014},
  volume  = {2014},
  number  = {239}
}

@article{Wilkinson2016,
  author  = {Wilkinson, Mark D. and Dumontier, Michel and Aalbersberg, IJsbrand Jan and Appleton, Gabrielle and Axton, Myles and others},
  title   = {The FAIR Guiding Principles for scientific data management and stewardship},
  journal = {Scientific Data},
  year    = {2016},
  volume  = {3},
  pages   = {160018},
  doi     = {10.1038/sdata.2016.18}
}

@misc{MappingInequality,
  author = {{Digital Scholarship Lab, University of Richmond}},
  title  = {Mapping Inequality: Redlining in New Deal America},
  year   = {2025},
  url    = {https://dsl.richmond.edu/panorama/redlining/}
}

@inproceedings{Jordahl2020,
  author    = {Jordahl, Kelsey and Van den Bossche, Joris and Fleischmann, Martin and McBride, James and Wasserman, Jacob and Gerard, Jeffrey and Badaracco, Adrian Gil and Snow, Alan D. and Tratner, Matthew and Perry, Matthew and et al.},
  title     = {GeoPandas: Python tools for geographic data},
  booktitle = {Proceedings of the 9th Python in Science Conference (SciPy 2020)},
  year      = {2020},
  editor    = {Huff, Katy and Bergstra, James},
  pages     = {120--132},
  doi       = {10.25080/Majora-342d178e-00e},
  publisher = {SciPy}
}

@misc{EPA2023,
  title = {{Environmental Protection Agency Air Quality System}},
  author = {{Environmental Protection Agency}},
  year = {2023},
  url = {https://www.epa.gov/air-data},
  note = {Accessed: 2023-06-15}
}

\end{document}